\documentclass[preprint]{ptephy_v1}
\usepackage{dcolumn}

\preprintnumber{KEK-CP-374}

\begin{document}

\title{
  Reconstruction of smeared spectral function from Euclidean
  correlation functions
}

\author[1]{Gabriela Bailas}
\author[1,2]{Shoji Hashimoto}
\author[2]{Tsutomu Ishikawa}
\affil[1]{Theory Center, High Energy Accelerator Research Organization (KEK), 
  Tsukuba 305-0801, Japan}
\affil[2]{School of High Energy Accelerator Science,
  The Graduate University for Advanced Studies (SOKENDAI),
  Tsukuba 305-0801, Japan}

\begin{abstract}
  We propose a method to reconstruct smeared spectral functions
  from two-point correlation functions measured on the Euclidean
  lattice.
  Arbitrary smearing function can be considered as far as it is smooth
  enough to allow an approximation using Chebyshev polynomials.
  We test the method with numerical lattice data of Charmonium
  correlators. 
  The method provides a framework to compare lattice calculation with
  experimental data including excited state contributions without
  assuming quark-hadron duality.
\end{abstract}

\date{\today}
\maketitle

\section{Introduction}
Reconstruction of hadron spectral function from Euclidean correlation
functions is a notoriously difficult problem.
In lattice quantum chromodynamics (LQCD) computations, which have so
far been the only practical method to calculate non-perturbative
quantities with errors under control, physical quantities are extracted
from $n$-point correlation functions obtained on an Euclidean lattice.
It means that all momenta inserted are space-like, so that physical
amplitudes, especially those for on-shell particles or resonances,
have to be read off from this unphysical setup.
The ground state contribution can be obtained relatively easily by
measuring an exponential fall-off of the correlator at long distances,
while excited states are much harder to identify because the
exponential function of the form $\exp(-Et)$ with an energy $E$ and a
time separation $t$ is numerically very similar for different $E$'s
especially when many energy levels are close to each other as in the
experimental situation.
Typically, then, only one or even none of the excited states can be
identified. 

Still, because the spectral function is of phenomenological interest
for various applications, several groups developed methods to extract
it with some extra pieces of information.
The maximum entropy method \cite{Nakahara:1999vy,Asakawa:2000tr} is
one of such attempts, where one assumes that unknown functions (or its
parameters) are statistically equally distributed within the model
space and tries to determine the most ``likely'' function.
A slightly different statistical approach based on Bayesian statistics
was also proposed \cite{Burnier:2013nla}.
Unfortunately, they are not free from uncertainties since the
{\it statistical distribution} of the spectral function does not
really have theoretical basis.
Similar problem would also remain in a recent attempt to use machine
learning to reconstruct the spectral function \cite{Kades:2019wtd}.

Another attempt to approach the problem is the use of the
Backus-Gilbert method, which is a deterministic method to obtain a
smeared spectral function \cite{Hansen:2017mnd}.
The smearing kernel is automatically determined by the data, requiring
that the width of the smearing is minimized.
In practice, one has to relax the minimization by adding an extra term
to the function to be minimized in order to avoid numerical instability.
The smearing kernel is therefore unknown until one actually performs
such analysis.
There is also a proposal to arrange the Backus-Gilbert
method such that the smearing kernel obtained becomes close to what
one inputs \cite{Hansen:2019idp}.
Several such methods have been tested with mock data led from a known
spectral function \cite{Tripolt:2018xeo}, and there is no obvious best
solution found so far.

We propose an alternative method to reconstruct the spectral function
with smearing specified by arbitrary kernels.
The Chebyshev polynomials are introduced to approximate the kernel
function, and a smeared spectral function under this approximation is
obtained from lattice data of temporal correlator.
The procedure is deterministic and the systematic error due to a
truncation of the Chebyshev polynomials can be estimated.
Limitation of the method comes from statistical error of the lattice
data, which prevents one from using high order polynomials.

The smeared spectral function can offer an intermediate quantity that
can be used to compare experimental data with non-perturbative
theoretical calculation provided by LQCD.
For instance, let us consider the $R$-ratio $R(s)$ defined for the
$e^+e^-\to q\bar{q}$ cross section as
$R(s)=\sigma_{e^+e^-\to q\bar{q}}(s)/\sigma_{e^+e^-\to\mu^+\mu^-}(s)$.
It cannot be directly compared with perturbative calculations in the
resonance region where perturbative expansion does not converge.
The LQCD calculation as it stands is not useful either, because it
can only calculate low-lying hadron spectrum and scattering phase
shift for specified final states, such as $\pi^+\pi^-$ or $K\bar{K}$,
but fully inclusive rate is unavailable.
Our method concerns how to extract the information for inclusive
processes, such as $q\bar{q}$, from lattice results of hadron
correlators without specifying any final states.
The $R$-ratio as a function of invariant mass of the final states
is not directly accessible in our method, but the function that is
smeared with some kernel can be related to quantities calculable on
the lattice. 

The smeared spectral function was considered in the early days of
perturbative QCD by Poggio, Quinn and Weinberg \cite{Poggio:1975af}.
They considered a smearing of the form
\begin{equation}
  \label{eq:smearing}
  \bar{R}(s,\Delta_s)=\frac{\Delta_s}{\pi}\int_0^\infty ds'
  \frac{R(s')}{(s'-s)^2+\Delta_s^2},
\end{equation}
whose kernel approaches a delta function $\delta(s-s')$ in the limit
of the width $\Delta_s\to 0$.
The $R$-ratio (or the spectral function) can be related to the vacuum
polarization function through the optical theorem
$R(s)=(1/\pi){\rm Im}\Pi(s)$
and the dispersion relation\footnote{In practice, one needs to use
  the subtracted version to avoid ultraviolet divergences.}
\begin{equation}
  \label{eq:dispersion}
  \Pi(q^2)=\frac{1}{\pi}\int_0^\infty ds \frac{{\rm Im}\Pi(s)}{s-q^2}.
\end{equation}
The smeared spectrum can then be written using the vacuum polarization
function at complex values of $q^2$:
\begin{equation}
  2i\bar{R}(s,\Delta_s)=\Pi(s+i\Delta_s)-\Pi(s-i\Delta_s).  
\end{equation}
The main observation was that, unlike the imaginary part
${\rm Im}\Pi(s)$ evaluated on the cut,
one can avoid non-perturbative kinematical region and calculate
$\Pi(s+i\Delta_s)$ in perturbation theory as long as the smearing
range $\Delta_s$ is large enough.
This is the argument behind the quark-hadron duality.
The width parameter $\Delta_s$ is typically chosen larger than the QCD
scale $\Lambda_{\rm QCD}$, but there is no {\it a priori} criteria of
how large $\Delta_s$ should be for perturbation theory to work to a
desired accuracy.
One can also consider a Gaussian smearing instead of
(\ref{eq:smearing}), and arrive at the same conclusion
\cite{Bertlmann:1984ih}. 

In many applications of perturbative QCD, such as deep inelastic
scattering or inclusive hadron decays, the smearing is not as
transparent as in this example.
Some smearing over kinematical variables is involved depending on the
setup of the problems, and the question of how much smearing is
introduced is more obscure.
Yet, one usually assumes that perturbation theory works; systematic
error due to this duality assumption is unknown.

Another commonly used form of smearing is the Laplace transform
\begin{equation}
  \label{eq:Laplace}
  \tilde{\Pi}(M^2)=\frac{1}{M^2}\int_0^\infty\! ds\,
  \left[\frac{1}{\pi} {\rm Im}\Pi(s)\right]
  e^{-s/M^2},
\end{equation}
which is related to the Borel transform involved in QCD sum rule
calculations \cite{Shifman:1978bx}.
Since $\tilde{\Pi}(M^2)$ can be written using the vacuum polarization
function $\Pi(Q^2)$ in the space-like domain, the non-perturbative
kinematical region is avoided.
The effective range of smearing is controlled by the parameter $M^2$.

Of course, one can view the dispersion relation
(\ref{eq:dispersion}) for a space-like value of $q^2=-Q^2$:
\begin{equation}
  \Pi(Q^2)=\frac{1}{\pi}\int_0^\infty ds\,\frac{{\rm Im}\Pi(s)}{s+Q^2},
\end{equation}
or its subtracted version
\begin{equation}
  \Pi(Q^2)-\Pi(0)=
  -\frac{Q^2}{\pi}\int_0^\infty ds\,\frac{{\rm Im}\Pi(s)}{s(s+Q^2)},
\end{equation}
as a sort of smearing.
The range of smearing is effectively infinity as the weight function
decreases only by a power of $s$.
This is another way of keeping away from the resonance region, and
perturbation theory is expected to be applicable.
One still needs to include power corrections using the operator
product expansion, which involves unknown parameters (or condensates);
LQCD calculation is desirable to eliminate such uncertainties.

We develop a formalism of LQCD calculation that allows us to compute
the quantities mentioned above,
{\it i.e.} those obtained by applying some smearing on the
spectral function.
The smearings are designed to escape from the resonance region to some
extent, so that perturbation theory is applicable, but some
uncertainty still remains as mentioned above.
By using LQCD, on the other hand, fully non-perturbative calculation
can be achieved and no remnant uncertainty due to the singularities
nor the power corrections remains.
In other words, one can entirely avoid the assumption of quark-hadron
duality. 

Through this method, the comparison between experimental data and
lattice calculation would provide a theoretically clean test of QCD.
It can also provide a testing ground for the perturbative QCD
analysis including the operator product expansion against the fully
non-perturbative lattice calculation.

This paper is organized as follows.
In Section~\ref{sec:spectral_function} we introduce the method to
reconstruct the smeared spectral function.
It includes an approximation using the Chebyshev polynomials, whose
performance is demonstrated for several cases with toy examples in
Section~\ref{sec:cheb_approx}.
Then, the method is tested with actual lattice data of Charmonium
correlator in Section~\ref{sec:charmonium}.
The discussion section (Section~\ref{sec:discussions})
lists possible applications of the methods including phenomenological
ones as well as those of theoretical interests related to the QCD sum
rule.
Our conclusions are given in Section~\ref{sec:conclusions}.

\section{Reconstruction of smeared spectral function}
\label{sec:spectral_function}
We are interested in the spectral function
\begin{equation}
  \label{eq:spectral_func}
  \bar\rho(\omega)=\frac{\langle\psi|\delta(\hat{H}-\omega)
    |\psi\rangle}{\langle\psi|\psi\rangle}
\end{equation}
for some state $|\psi\rangle$.
It does not have to be an eigenstate of Hamiltonian $\hat{H}$, but can be 
created by applying some operator on the vacuum, {\it e.g.} 
$\sum_xJ_\mu(x)|0\rangle$ for the case of $e^+e^-\to q\bar{q}$.
Here $J_\mu(x)$ is the electromagnetic current and the sum over $x$
gives the (spatial) zero-momentum projection.
An extension to the case of different initial and final states should
be possible.
The spectral function $\bar\rho(\omega)$ in (\ref{eq:spectral_func})
is normalized such that it becomes unity when integrated over all
possible energy $\omega$.

On the lattice, we calculate the temporal correlation function
\begin{equation}
  \label{eq:correlator}
  \bar{C}(t) = \frac{\langle\psi|e^{-\hat{H}t}|\psi\rangle}{
    \langle\psi|\psi\rangle},
\end{equation}
which is normalized to one at zero time separation.
It can be rewritten using the spectral function as
\begin{equation}
  \label{eq:rhobar}
  \bar{C}(t) = 
  \frac{\langle\psi| \int_0^\infty\! d\omega\,
    \delta(\hat{H}-\omega) e^{-\omega t} |\psi\rangle}{
    \langle\psi|\psi\rangle} =
  \int_0^\infty\! d\omega\,\bar\rho(\omega)e^{-\omega t},
\end{equation}
thus the conventional spectral decomposition of a
correlator.

In practice, we can take the state $|\psi\rangle$ as
\begin{equation}
  |\psi\rangle = e^{-\hat{H}t_0} \sum_x V_\mu|0\rangle  
\end{equation}
with some (small, but non-zero) time separation $t_0$ 
in order to avoid any potential divergence due to a contact term when
evaluating $\langle\psi|\psi\rangle$.
For instance, by taking $t_0=1$ in the lattice unit,
we can identify (\ref{eq:correlator}) as $C(t+2)/C(2)$ for a
vector correlator $C(t)=\langle 0|V_\mu(t)V_\mu(0)|0\rangle$, where 
$\mu$ stands for a spatial direction and not summed over.

We note that the Hamiltonian $\hat{H}$ in (\ref{eq:correlator}) is not 
explicitly written in lattice QCD simulations,  
but we assume that it exists so that the time evolution is written by
a transfer matrix $\hat{z}=e^{-\hat{H}}$.
We also assume that the eigenvalues of $\hat{H}$ are non-negative and
equivalently that the eigenvalues of $\hat{z}$ are constrained to lie
between 0 and 1.
Strictly speaking, many lattice actions currently used in numerical
simulations do not satisfy the {\it reflection positivity}, which is a
necessary condition for a Hermitian Hamiltonian to exist.
Any problem due to the violation of the reflection positivity is
expected to disappear in the continuum limit.
Our assumption is, therefore, that our lattice calculations are
sufficiently close to the continuum limit.
Using the transfer matrix $\hat{z}$,
the correlator in (\ref{eq:correlator}) is simply written as
$\bar{C}(t)=\langle\psi|\hat{z}^t|\psi\rangle/\langle\psi|\psi\rangle$.

For the vacuum polarization function due to electromagnetic currents,
\begin{equation}
  \Pi_{\mu\nu}(q)=(q_\mu q_\nu-q^2\delta_{\mu\nu})\Pi(q^2)
  = \int\!d^4x\,e^{iq\cdot x}\langle 0|J_\mu(x)J_\nu(0)|0\rangle,
\end{equation}
the spectral function is often defined as
$\rho(s)=(1/\pi){\rm Im}\Pi(s)$.
The Euclidean correlator is then expressed as
\begin{equation}
  C(t)=\int_0^\infty d\omega\,\omega^2\rho(\omega^2) e^{-\omega t}.
\end{equation}
(See \cite{Bernecker:2011gh}, for instance.)
This is slightly different from (\ref{eq:correlator}), but can be
related by redefining the spectral function.
Namely, we define $\bar{C}(t)$ as
$\bar{C}(t)\equiv C(t+2t_0)/C(2t_0)$, so that the spectral function
$\bar{\rho}(\omega)$ in (\ref{eq:rhobar}) is
$\bar{\rho}(\omega)=(1/C(2t_0))\omega^2\rho(\omega^2)e^{-2\omega t_0}$.
We note that the normalization factor $C(2t_0)$ is explicitly
calculable on the lattice.

Now we define a smeared spectral function $\bar\rho_\Delta(\omega)$
for a smearing kernel $S_\Delta(\omega,\omega')$ as
\begin{eqnarray}
  \bar\rho_\Delta(\omega) &=& \int_0^\infty d\omega' S_\Delta(\omega,\omega')
                         \bar\rho(\omega')
  \\
  &=&
  \frac{\langle\psi| \int_0^\infty d\omega'
  S_\Delta(\omega,\omega')
  \delta(\hat{H}-\omega') |\psi\rangle}{\langle\psi|\psi\rangle}
  \\
  &=& \frac{\langle\psi| S_\Delta(\omega,\hat{H})|\psi\rangle}{
      \langle\psi|\psi\rangle}.
\end{eqnarray}
Then, the matrix element to be evaluated is
$\langle\psi|S_\Delta(\omega,\hat{H})|\psi\rangle$.
The form of the smearing kernel $S_\Delta(\omega,\omega')$ is
arbitrary; we can consider the choices discussed in the previous
section. 
In the following, to be explicit, let us assume a specific form for
$S_\Delta(\omega,\omega')$ as
\begin{equation}
  S_\Delta(\omega,\omega')=\frac{1}{\pi}
  \frac{2\Delta}{(\omega-\omega')^2+\Delta^2},
  \label{eq:smearing}
\end{equation}
where $\Delta$ represents a range of smearing.

We consider a polynomial approximation of $S_\Delta(\omega,\hat{H})$ of
the form
\begin{equation}
  S_\Delta(\omega,\hat{H}) \simeq \frac{c_0(\omega)}{2} +
  \sum_{j=1}^N  c_j(\omega) T_j(\hat{z}),
  \label{eq:cheb_approx}
\end{equation}
where $T_j(x)$ stands for the Chebyshev polynomial and the sum is up
to its maximal order $N$.
The first few terms of the Chebyshev polynomials are
$T_0(x)=1$, $T_1(x)=x$, $T_2(x)=2x^2-1$, $T_3(x)=4x^3-3x$, $\cdots$,
and one can use the recursion relation
$T_{j+1}(x)=2xT_j(x)-T_{j-1}(x)$ to construct the following ones.
According to the general formula of the Chebyshev approximation,
the coefficients $c_j(\omega)$ in (\ref{eq:cheb_approx}) can be
obtained as
\begin{equation}
  c_j(\omega)=\frac{2}{\pi}\int_{-1}^1 \frac{dx}{\sqrt{1-x^2}} f_\omega(x) T_j(x),
  \label{eq:cheb_coeff}
\end{equation}
where the function $f_\omega(x)$ is written as
\begin{equation}
  f_\omega(x) = \left\{
    \begin{array}{ll}
      \displaystyle
      \frac{1}{\pi}\frac{2\Delta}{(\omega + \ln x)^2+\Delta^2} & (0<x\le 1),
      \\
      0 & (-1\le x\le 0).
    \end{array}
  \right.
\end{equation}
Here, $x$ corresponds to eigenvalues of $\hat{z}=e^{-\hat{H}}$, 
so that the function $f_\omega(x)$ represents the smearing function
(\ref{eq:smearing}) with $\omega'$ replaced by $\hat{H}$.
This standard formula for the Chebyshev approximation is
written for a function $f_\omega(x)$ defined in $[-1,1]$.
Here we use it only between $[0,1]$ and assume that $f_\omega(x)$
vanishes for $x\le 0$.
Technically, the numerical integral (\ref{eq:cheb_coeff}) becomes
unstable for large $j$'s due to a divergence of the integrand as 
$x\to 1$. 
Instead, one may use an alternative formula
\begin{equation}
  c_j(\omega)=\frac{2}{\pi}\int_0^\pi d\theta\,
  f_\omega(\cos\theta) \cos(j\theta),
\end{equation}
evaluation of which is more stable for large $j$'s.
For the range of $x$ between $[0,1]$, this integral is up to $\pi/2$.

To optimize the Chebyshev approximation, one can use a modified form
written in terms of the shifted Chebyshev polynomials
$T^*_n(x)\equiv T_n(2x-1)$, which is defined in
$0\le x\le 1$.
Its first few terms are 
$T_0^*(x)=1$, $T_1^*(x)=2x-1$, $T_2^*(x)=8x^2-8x+1$,
$T_3^*(x)=32x^3-48x^2+18x-1$, $\cdots$.
The corresponding formula for the coefficients appearing in the
Chebyshev approximation is 
\begin{equation}
  c_j^*(\omega)=\frac{2}{\pi}\int_0^\pi d\theta\,
  f_\omega\left(\frac{1+\cos\theta}{2}\right) \cos(j\theta).
\end{equation}
The approximation formula (\ref{eq:cheb_approx}) is unchanged
other than replacing $c_j(\omega)T_j(\hat{z})$ by
$c_j^*(\omega)T_j^*(\hat{z})$.
Since the range of $x$ is narrower, this series gives a better
approximation of the original function for a given order $N$.

Finally, remember that the matrix elements of the transfer matrix
$\hat{z}$ and its power $\hat{z}^t$ can be written as
$\bar{C}(t)=
\langle\psi|\hat{z}^t|\psi\rangle/\langle\psi|\psi\rangle$.
Then, we arrive at an expression for $\bar\rho_\Delta(\omega)$:
\begin{equation}
  \label{eq:rhoDelta_approx}
  \bar\rho_\Delta(\omega) \simeq \frac{c_0^*(\omega)}{2}
  +\sum_{j=1}^N c_j^*(\omega)\langle T_j^*(\hat{z})\rangle,
\end{equation}
where the last term $\langle T_j^*(\hat{z})\rangle$
may be constructed from the correlator $\bar{C}(t)$
by replacing the power of the transfer matrix $\hat{z}^t$ appearing in
$T_j^*(\hat{z})$ by $\bar{C}(t)=C(t+2t_0)/C(2t_0)$.
Therefore, the first few terms are obtained as
\begin{eqnarray}
  \label{eq:construct_Tj}
  \langle T_0^*(\hat{z})\rangle & = & 1,\nonumber\\
  \langle T_1^*(\hat{z})\rangle & = & 2\bar{C}(1)-1,\nonumber\\
  \langle T_2^*(\hat{z})\rangle & = & 8\bar{C}(2)-8\bar{C}(1)+1,\nonumber\\
  \langle T_3^*(\hat{z})\rangle & = & 32\bar{C}(3)-48\bar{C}(2)+18\bar{C}(1)-1,
  \\ & \vdots & \nonumber
\end{eqnarray}
which is a deterministic procedure.

The general expression (\ref{eq:rhoDelta_approx}) for an approximation
of $\bar\rho_\Delta(\omega)$ is valid for any smearing kernel and for
any value of $\omega$, as long as the coefficients $c_j^*(\omega)$ are 
calculated appropriately.
As it is well known, the Chebyshev approximation provides the
{\it best} approximation of any function defined in $0\le x\le 1$.
It is the best among any polynomials at a given order $N$ in the sense
that the minmax error, the maximum deviation from the true function in
the same range, is minimum; in order to achieve a better
approximation, one needs a higher polynomial order $N$.
Since the (shifted) Chebyshev polynomials $T_j^*(x)$ are oscillating
functions between 0 and 1, it is necessary to use larger $N$ in order
to better approximate detailed shape of the original spectral
function $\bar\rho(\omega)$ by narrowing the width $\Delta$ of the
smearing kernel. 
The approximation is demonstrated in the next section by taking a few 
examples.

The shifted Chebyshev approximation works only when the argument $x$
is in $0\le x\le 1$.
In our case, it corresponds to the condition that the eigenvalues of
$\hat{z}$ are in $[0,1]$, which should be satisfied because $\hat{z}$
is the transfer matrix $\hat{z}=e^{-\hat{H}}$.
For a given eigenvalue $z_i$ of $\hat{z}$, each polynomial $T_j^*(z_i)$
takes a value between $-1$ and $1$, and if the state $|\psi\rangle$ is
decomposed as $|\psi\rangle=\sum_ia_i|i\rangle$ with a normalization
$\sum_i|a_i|^2=\langle\psi|\psi\rangle$,
the individual polynomial becomes
$\langle\psi|T_j^*(\hat{z})|\psi\rangle
=\sum_i|a_i|^2 T_j^*(z_i)$,
which is bounded by $\pm\sum_i|a_i|^2=\pm\langle\psi|\psi\rangle$ so
that $\langle T_j^*(\hat{z})\rangle$ is bounded by $\pm 1$.
This provides a non-trivial constraint that must be satisfied by
the correlator $\bar{C}(t)$.

\section{Chebyshev polynomial approximation: examples}
\label{sec:cheb_approx}
\begin{figure}[tbp]
  \centering
  \includegraphics[width=8cm]{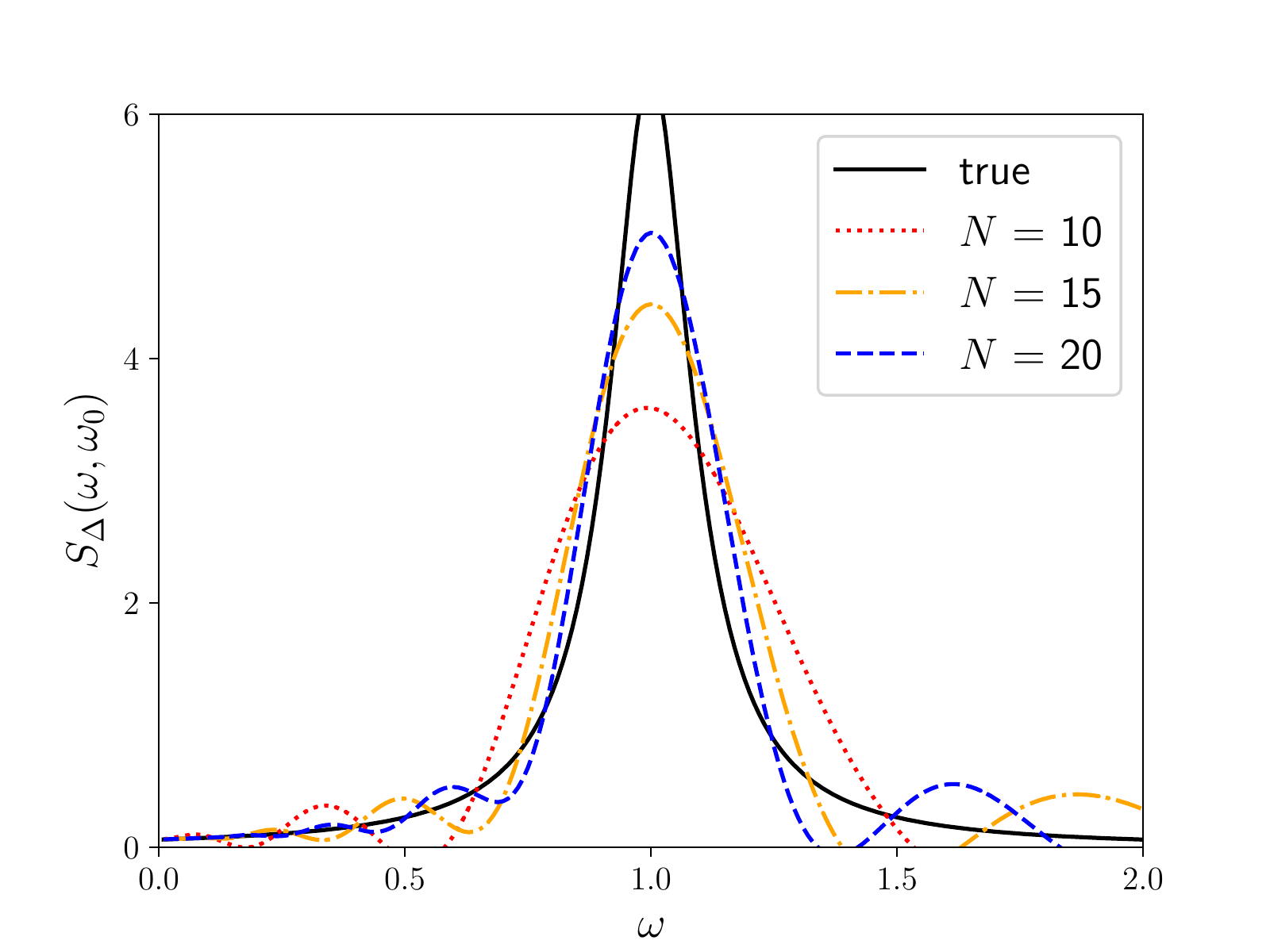}
  \includegraphics[width=8cm]{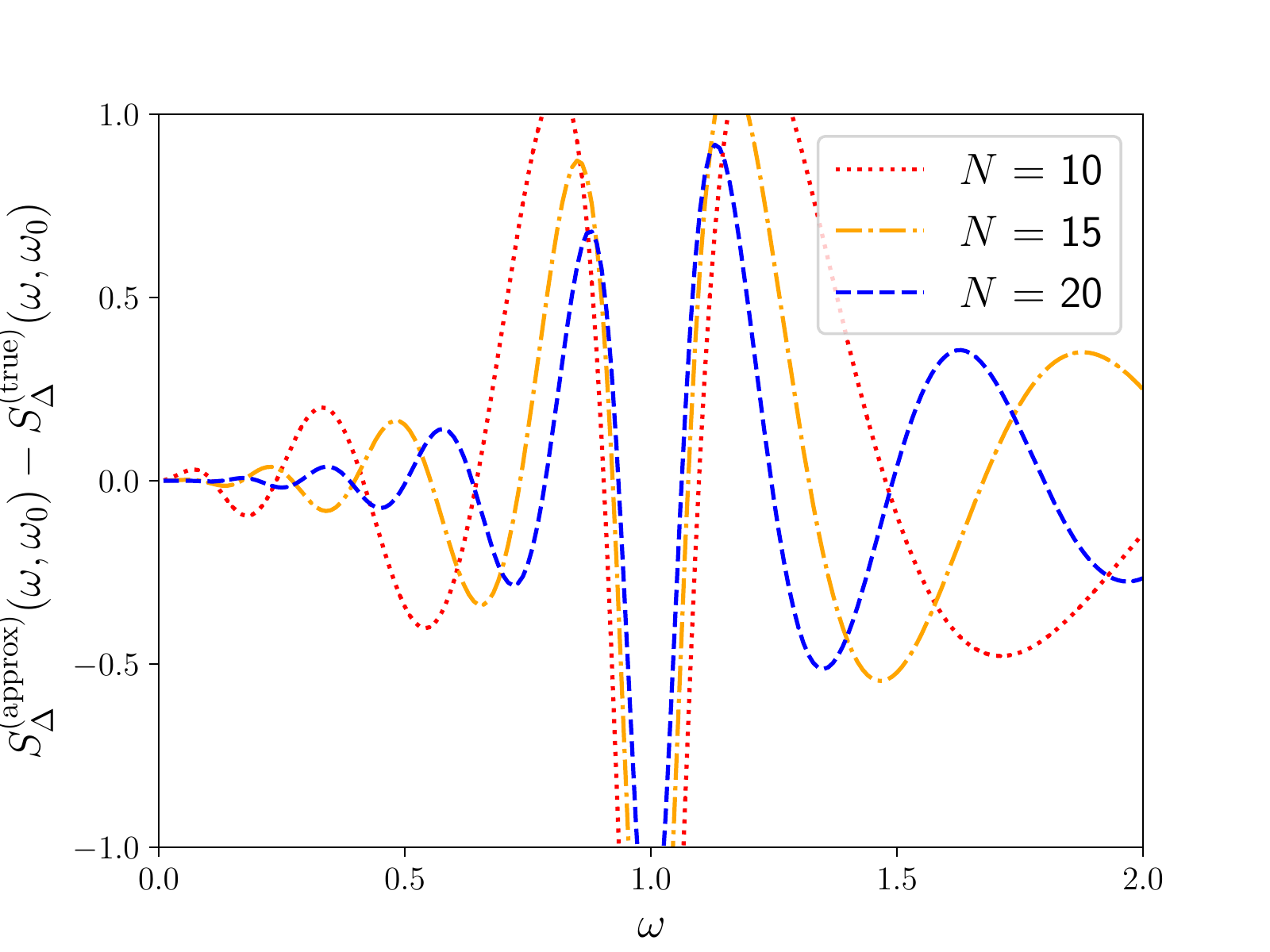}\\
  \includegraphics[width=8cm]{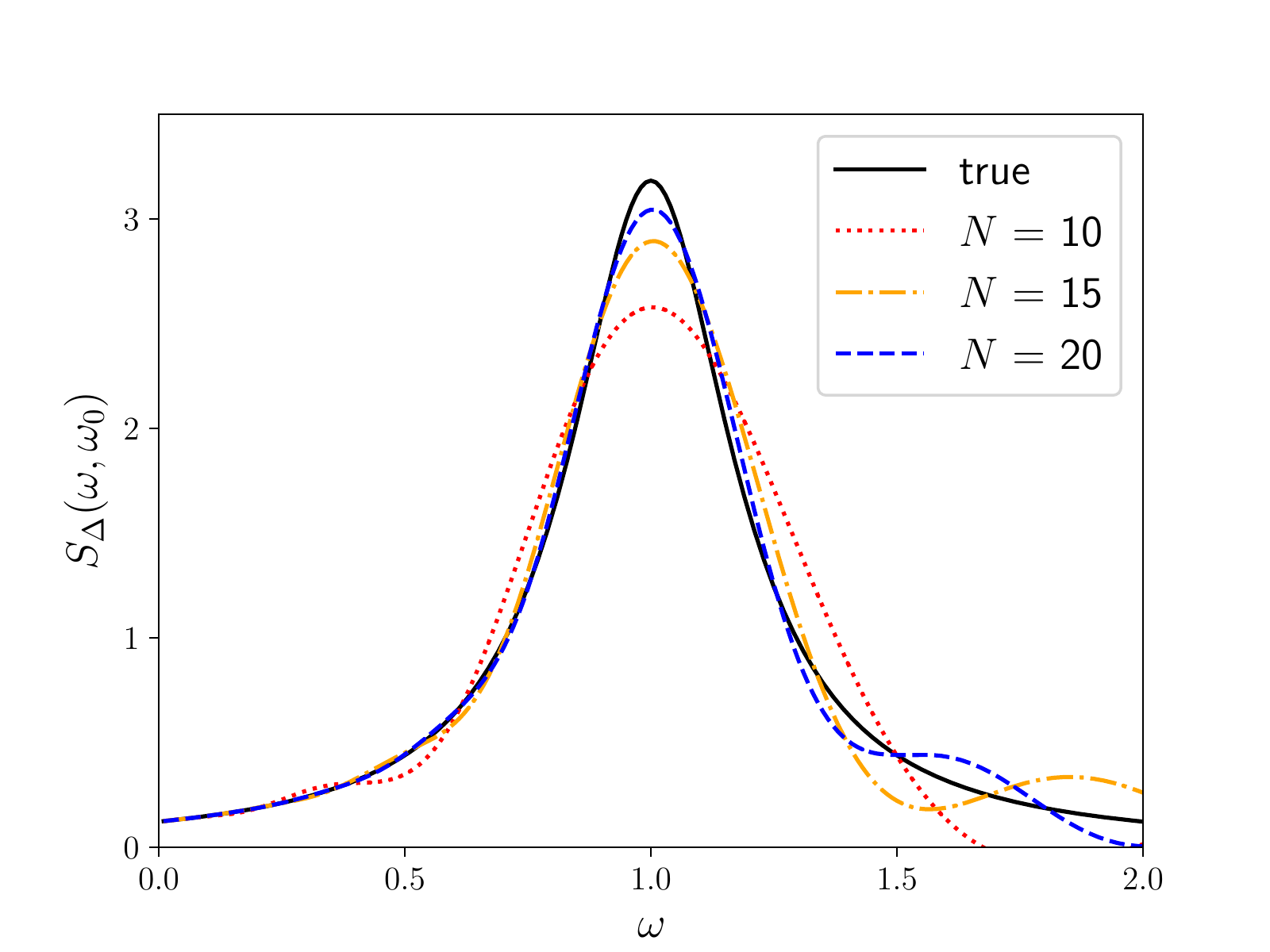}
  \includegraphics[width=8cm]{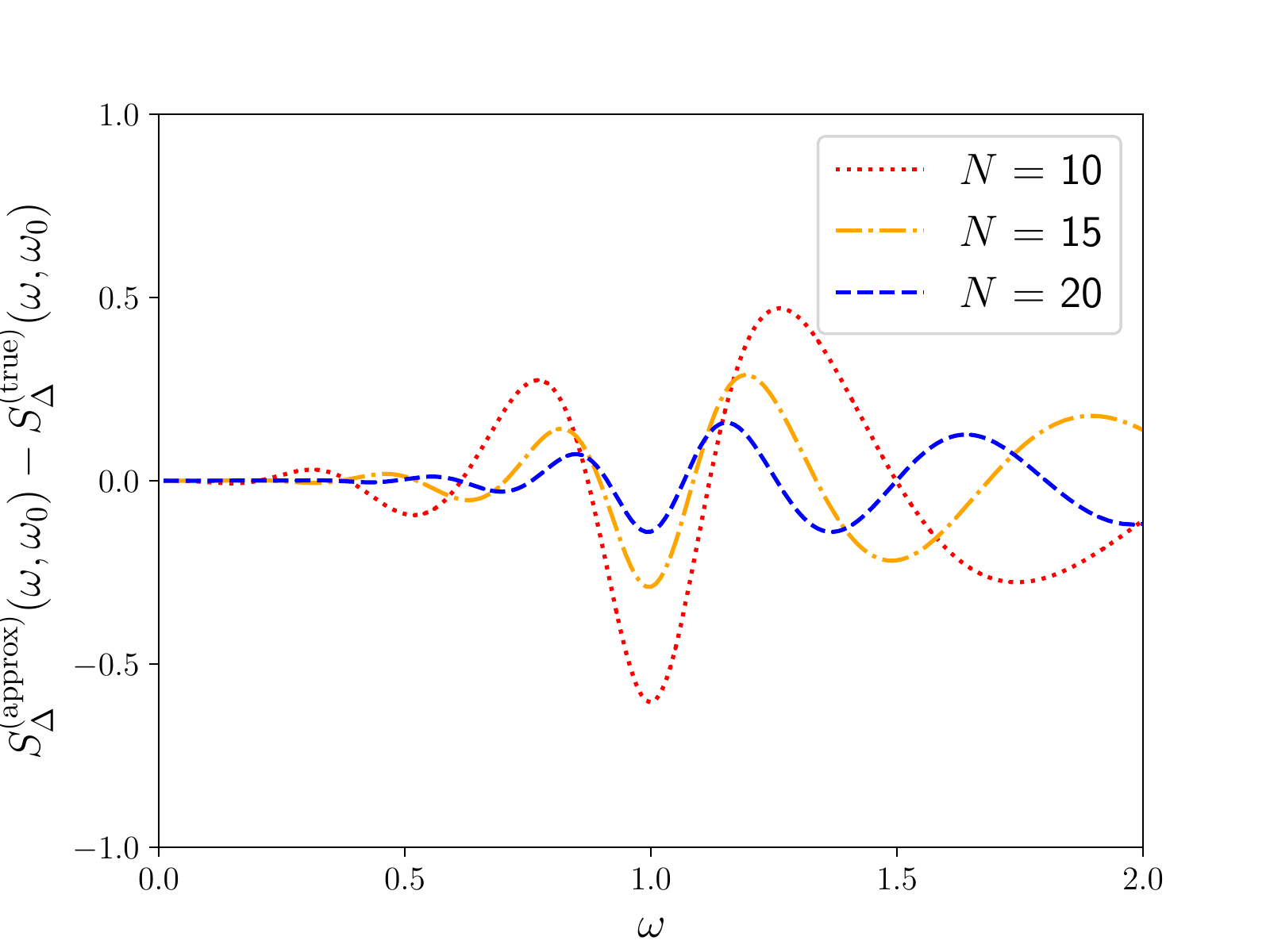}\\
  \includegraphics[width=8cm]{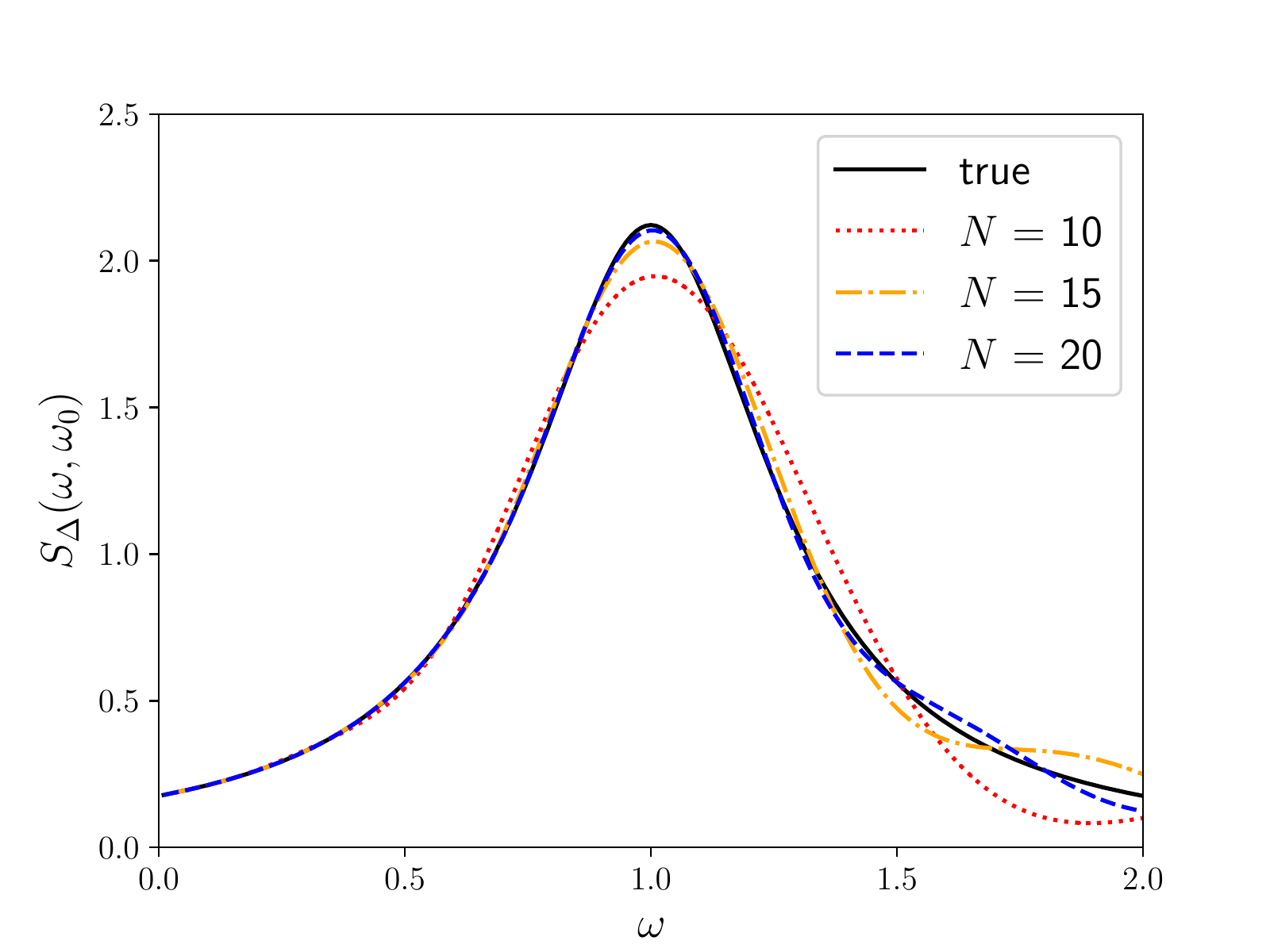}
  \includegraphics[width=8cm]{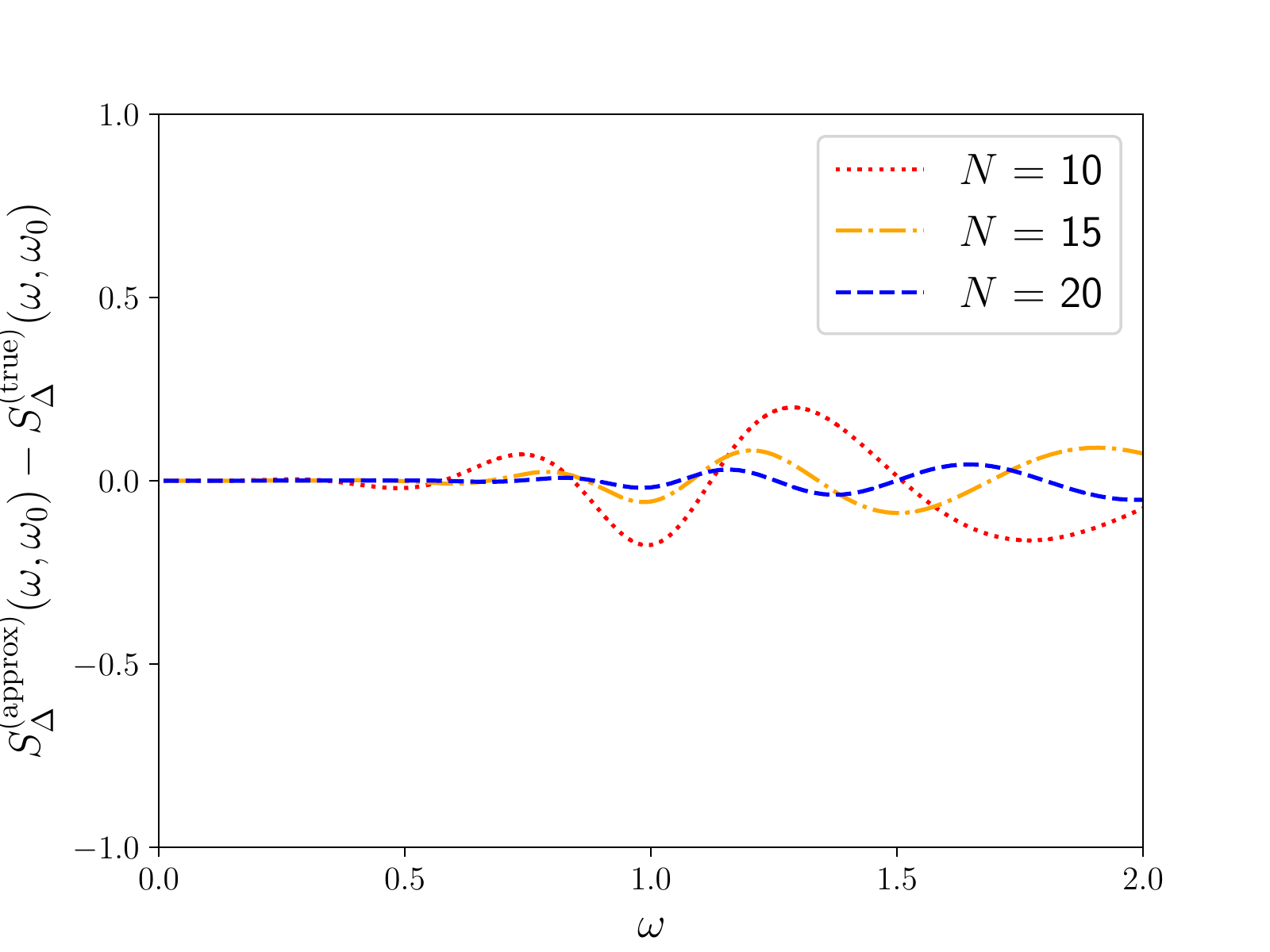}\\
  \caption{Left: Chebyshev approximation of the smearing kernel
    $S_\Delta(\omega,\omega_0)$ for $\omega_0=1$ and $\Delta$ = 0.1
    (top), 0.2 (middle) and 0.3 (bottom).
    Solid line is the true function, while dotted, dot-dashed and
    dashed lines are the approximations with $N$ = 10, 15 and 20,
    respectively.
    Right: Its error compared to the true function.
  }
  \label{fig:cheb_approx}
\end{figure}

First, we demonstrate how well the smearing kernel
$S_\Delta(\omega,\omega')$, Eq.~(\ref{eq:smearing}), is approximated by
the Chebyshev polynomials.
Setting $\omega'=\omega_0$ = 1, in some unit, say the lattice unit,
we draw a curve of $S_\Delta(\omega,\omega_0)$ in
Fig.~\ref{fig:cheb_approx} (left).
The Chebyshev approximation of the form (\ref{eq:cheb_approx}),
replacing $\hat{z}$ by $e^{-\omega_0}$ in this equation, is also
plotted for $N$ = 10 (dotted), 15 (dot-dashed) and 20 (dashed curve).
On the right panels, we plot the error of the approximation,
namely
a difference from the true function
$S_\Delta^{\rm (approx)}(\omega,\omega_0)-
 S_\Delta^{\rm (true)}(\omega,\omega_0)$.

From the plots one can confirm that the smearing function with a larger
width $\Delta$ = 0.3 is well approximated by a limited order of the
polynomials.
Namely, the polynomials up to order $N$ = 20 or even 15 give a nearly
perfect approximation; the deviation is at a few per cent level.
Apparently, the approximation becomes poorer when the function is
sharper, $\Delta$ = 0.2 or 0.1.
One needs higher order polynomials to achieve better approximation.
We limit ourselves to $N$ = 10--20, because these are the orders that
can be practically used for the analysis of lattice data, as we
discuss in the next section.

\begin{figure}[tbp]
  \centering
  \includegraphics[width=8cm]{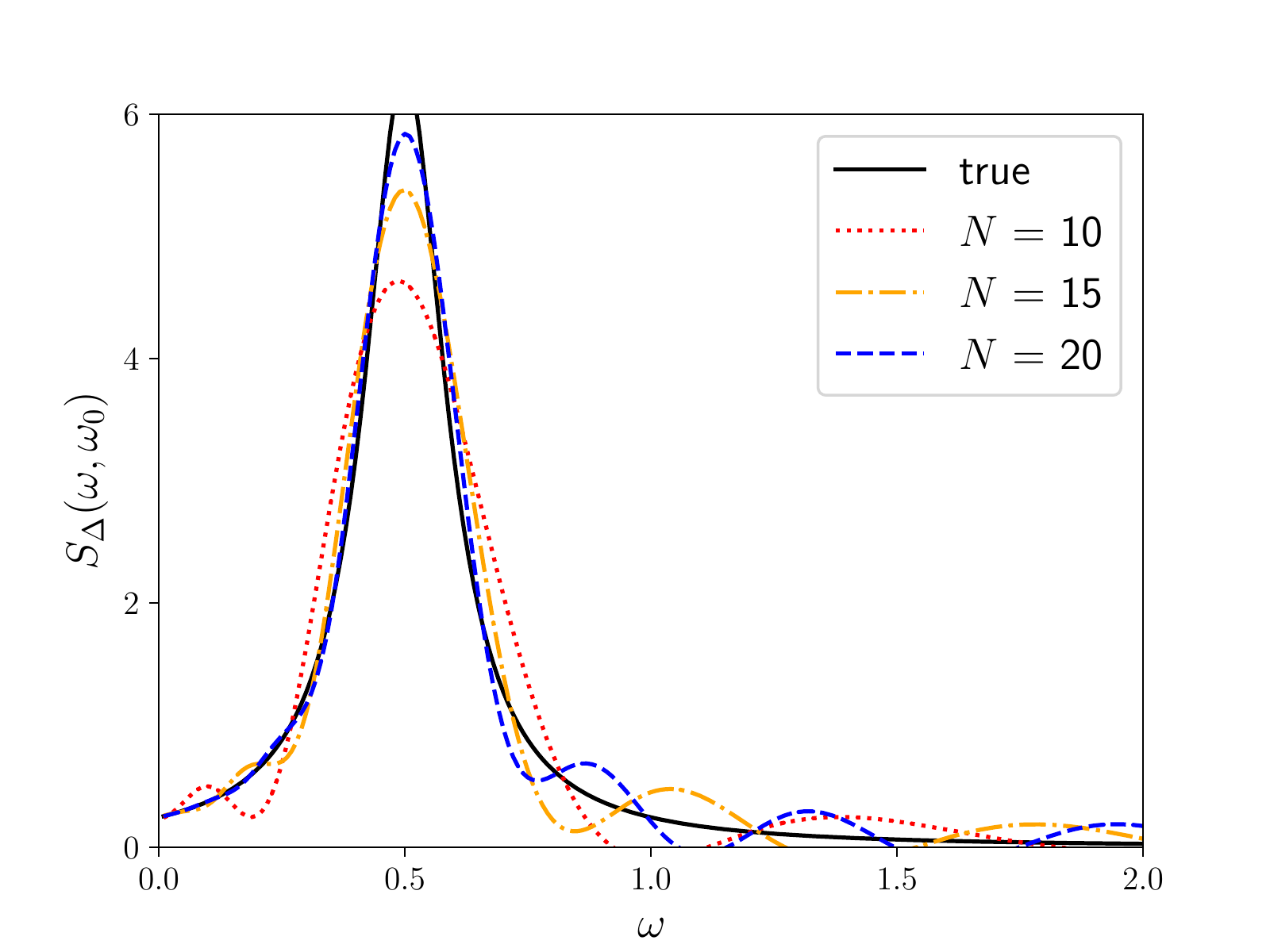}
  \includegraphics[width=8cm]{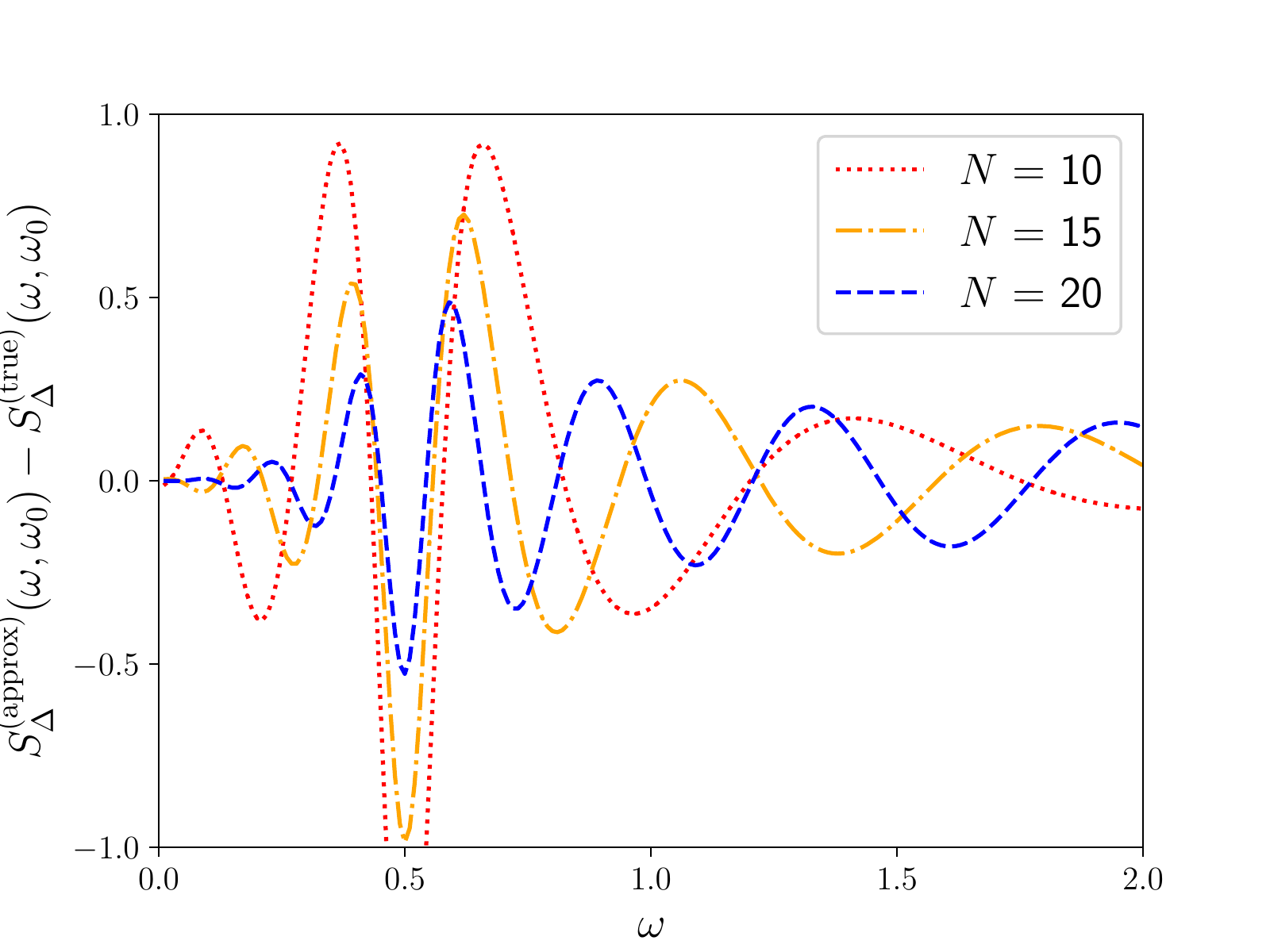}\\
  \includegraphics[width=8cm]{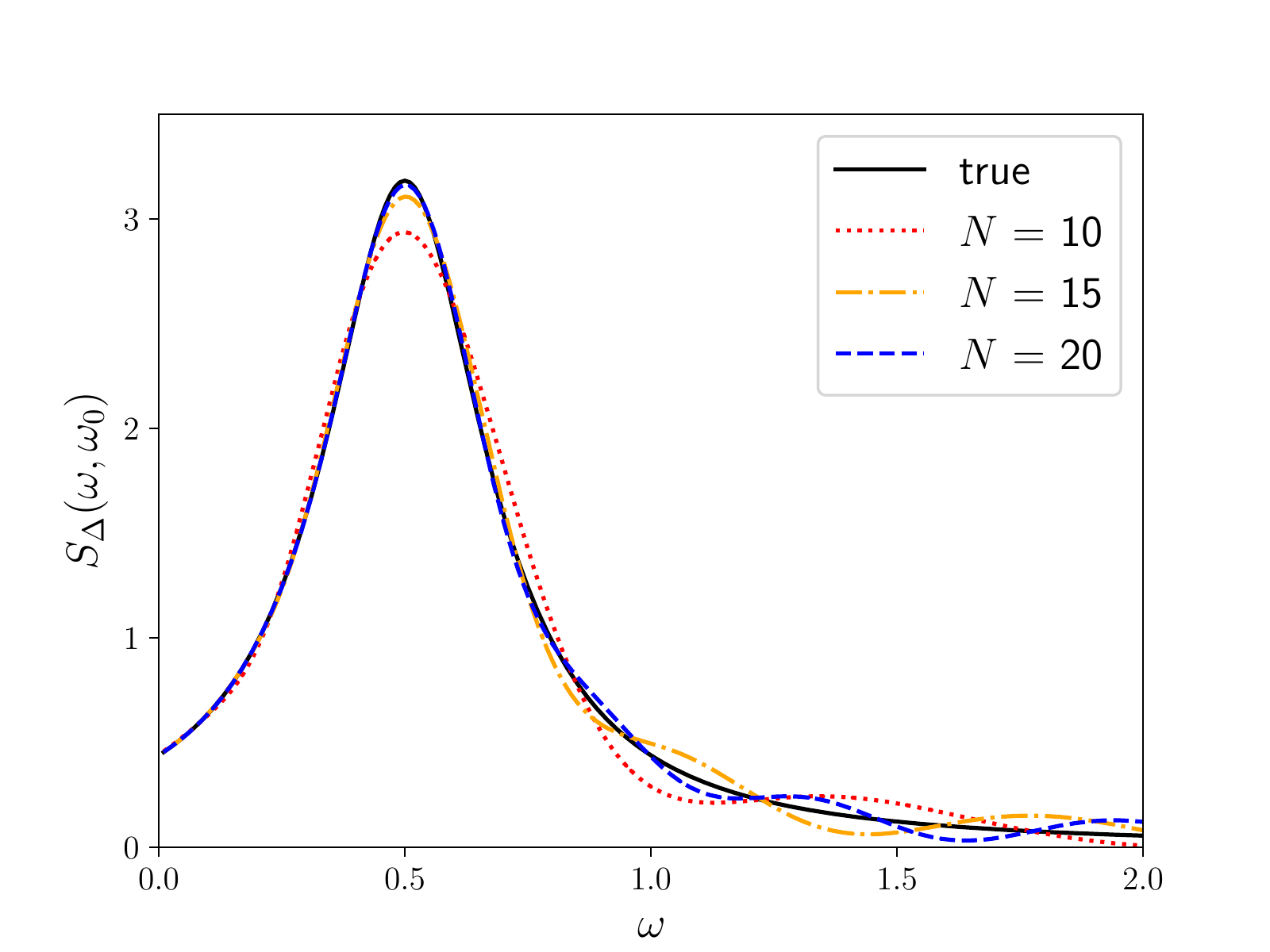}
  \includegraphics[width=8cm]{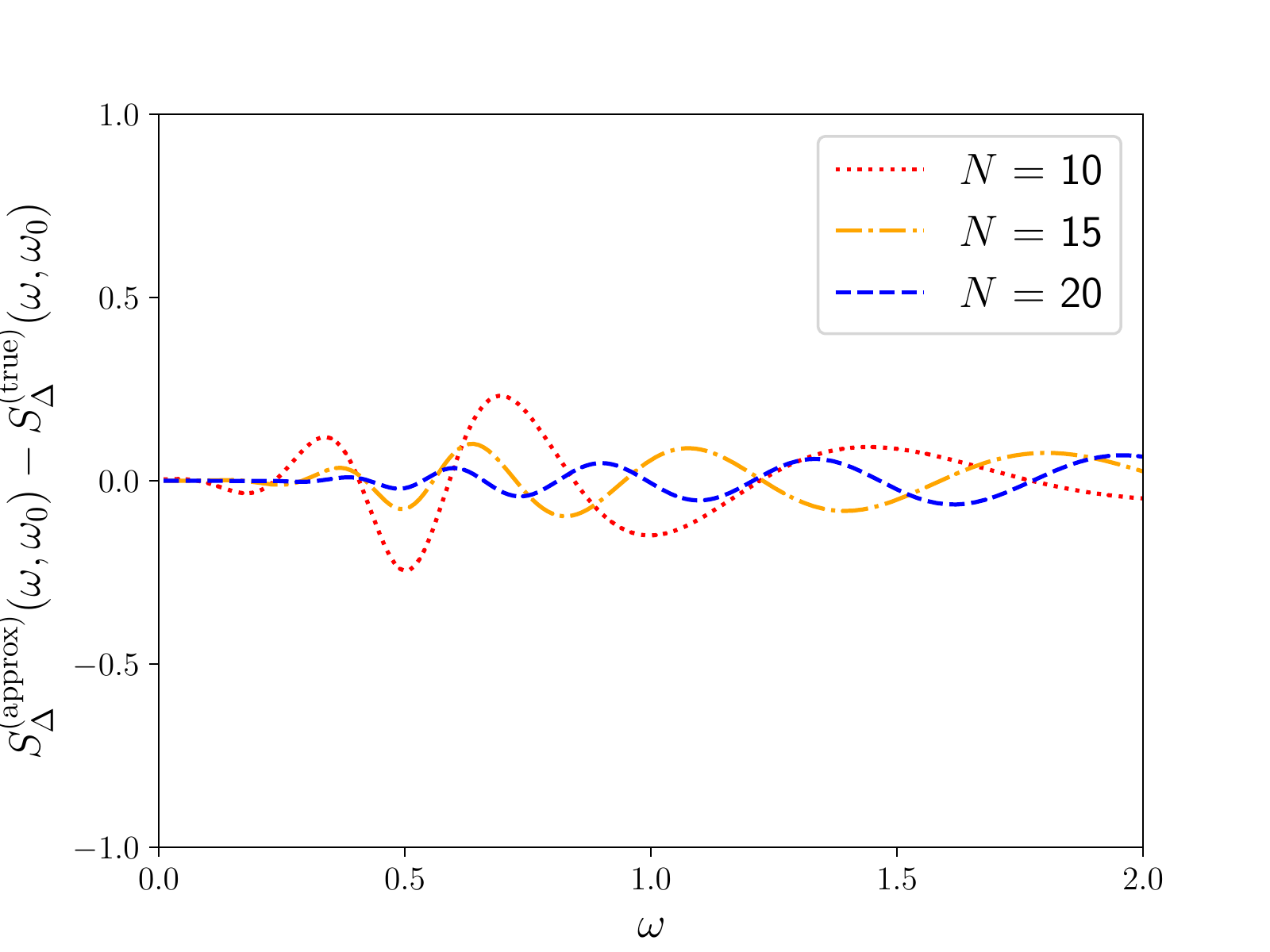}\\
  \includegraphics[width=8cm]{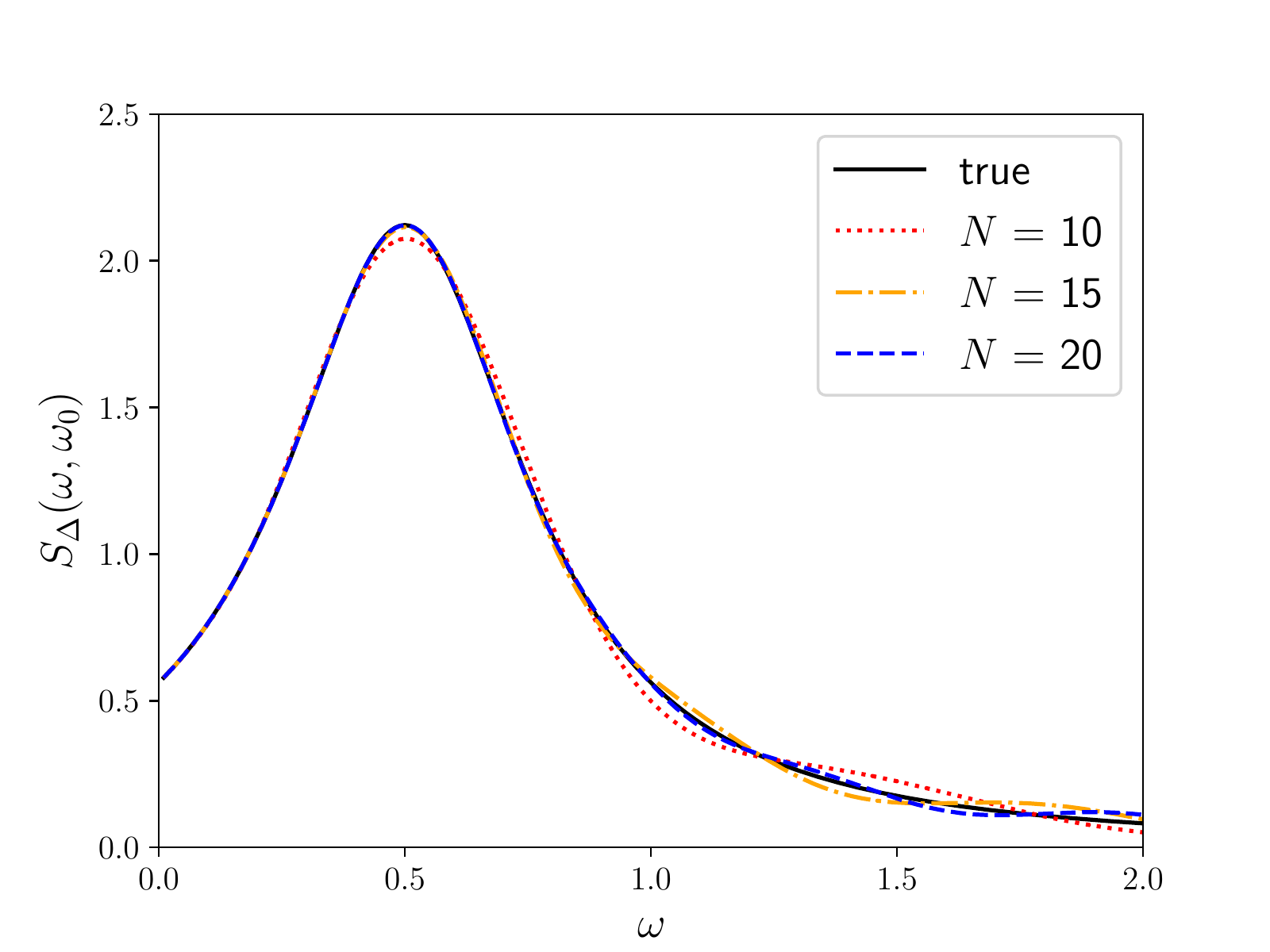}
  \includegraphics[width=8cm]{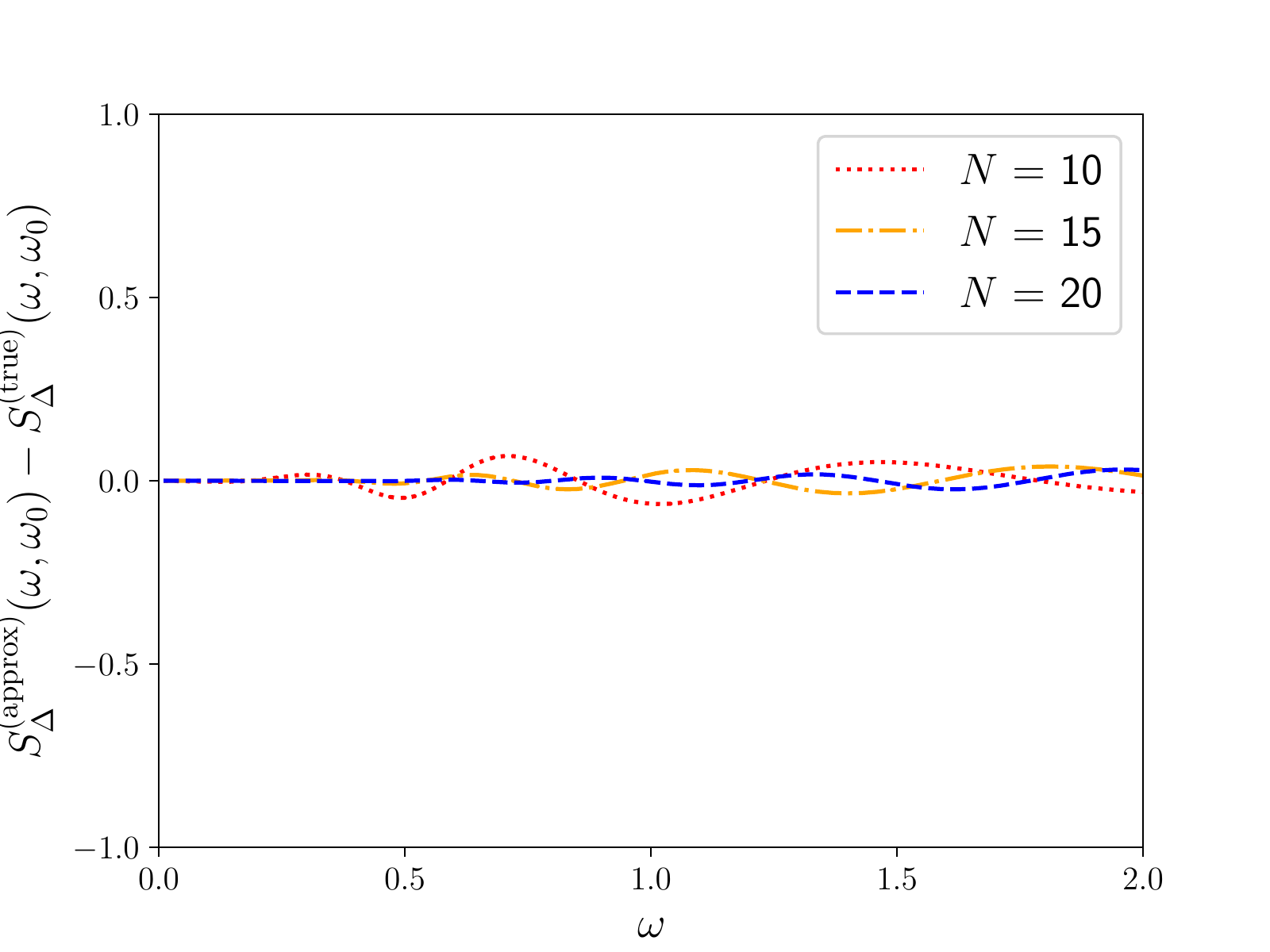}\\
  \caption{Same as Fig.~\ref{fig:cheb_approx} but for $\omega_0$ =
    0.5.
  }
  \label{fig:cheb_approx_w0=0.5}
\end{figure}

When the target energy $\omega_0$ is lower, $\omega_0$ = 0.5,
we observe a very similar pattern as shown in
Fig.~\ref{fig:cheb_approx_w0=0.5}.
An important difference is, however, that
the approximation is better than those for $\omega_0$ = 1.0,
as one can see by comparing the size of the error
$S_\Delta^{\rm (approx)}(\omega,\omega_0)-
 S_\Delta^{\rm (true)}(\omega,\omega_0)$.
This is probably because the approximation is constructed as a
function of $z=e^{-\omega}$, and the Chebyshev approximation works
uniformly between $z\in[0,1]$.
The range of $\omega\in[0.5,1.5]$, which is the central region for
$\omega_0=1$ is mapped onto $z\sim[0.22,0.61]$, while 
$\omega\in[0,1]$, for $\omega_0=0.5$ corresponds to $z\sim[0.37,1]$,
which stretches over a wider range and the Chebyshev approximation
works more efficiently.

\begin{figure}[tbp]
  \centering
  \includegraphics[width=8cm]{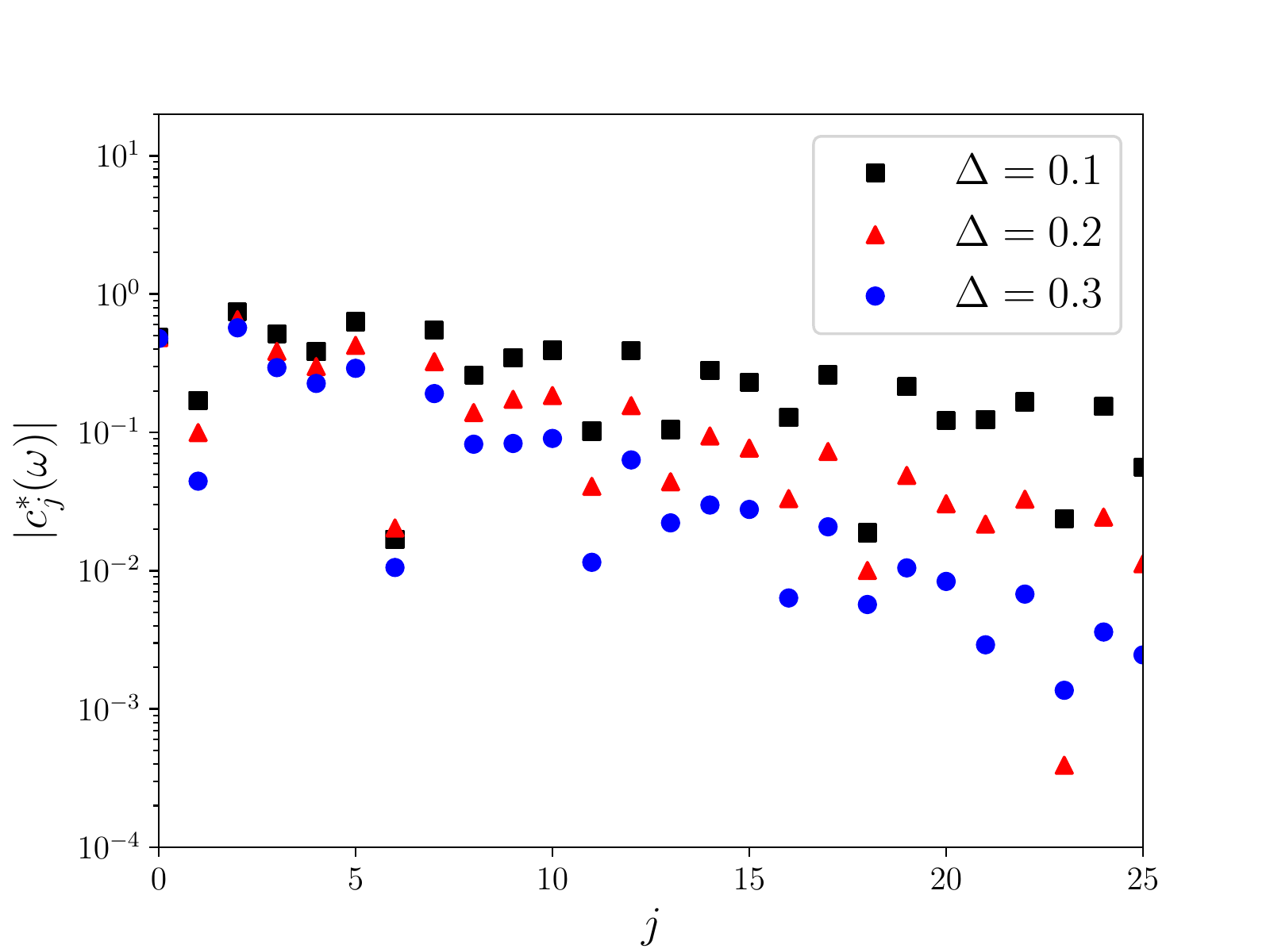}
  \includegraphics[width=8cm]{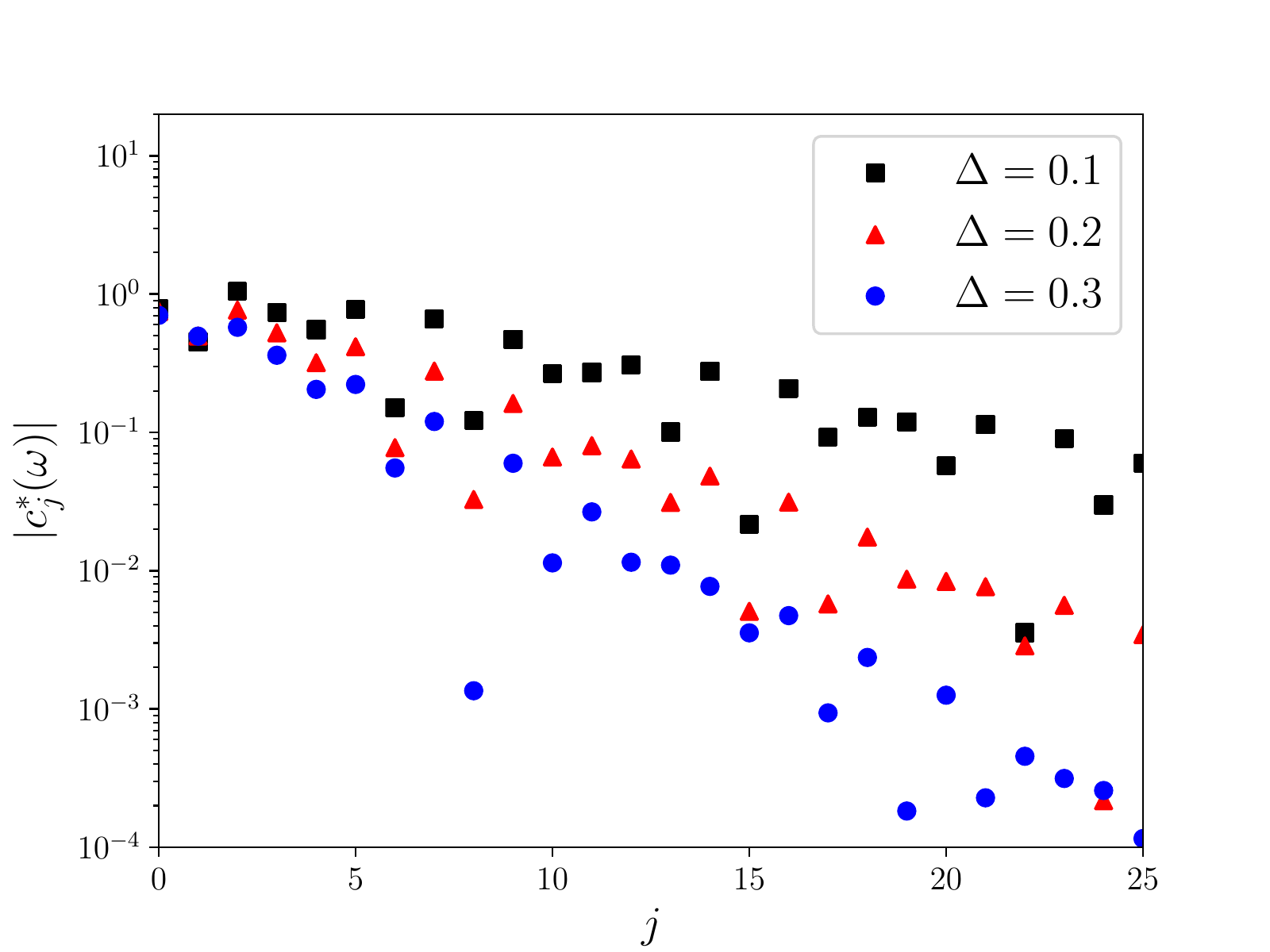}
  \caption{Coefficient of the Chebyshev approximation $c_j^*(\omega)$
    at $\omega=\omega_0$ plotted against $j$ for $\omega_0$ = 1.0
    (left) and 0.5 (right).
    Results for $\Delta$ = 0.1 (squares), 0.2 (triangles), and 0.3
    (circles). 
  }
  \label{fig:cheb_coeff}
\end{figure}

In order to see the rate of convergence of the Chebyshev
approximation, we plot the coefficients $c_j^*(\omega)$ as a function
of $j$ in Fig.~\ref{fig:cheb_coeff}.
As an example, the point $\omega=\omega_0$ is taken because this is
where the error is largest.
(It is not always the case especially when the approximation is
already good. See Figs.~\ref{fig:cheb_approx} and
\ref{fig:cheb_approx_w0=0.5}.)
One can see that the magnitude of the coefficient $|c_j^*(\omega)|$
decreases roughly exponentially as $j$.
When the approximation is better (larger $\Delta$), the decrease of
$|c_j^*(\omega)|$ is faster.
Since the Chebyshev polynomial $|\langle T_j^*(\hat{z})\rangle|$ is
bounded from above by 1, this shows (the upper limit of) the rate of
convergence.
The Calculation of $c_j^*(\omega)$ is numerically inexpensive, and one can
easily estimate the error of the approximation due to a truncation at
the order $N$ by $\pm|c_{N+1}^*(\omega)|$.

\begin{figure}[tbp]
  \centering
  \includegraphics[width=8cm]{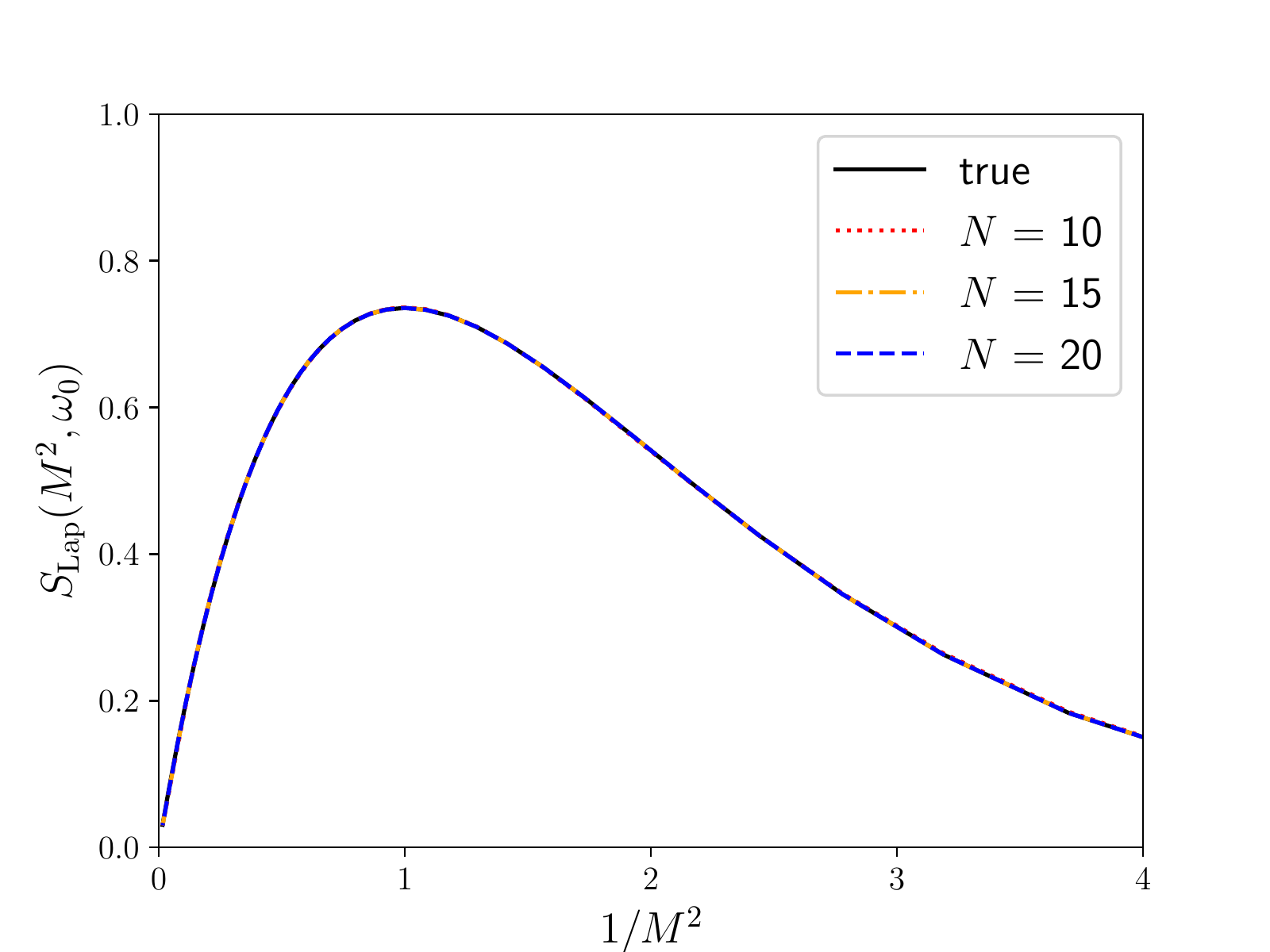}
  \includegraphics[width=8cm]{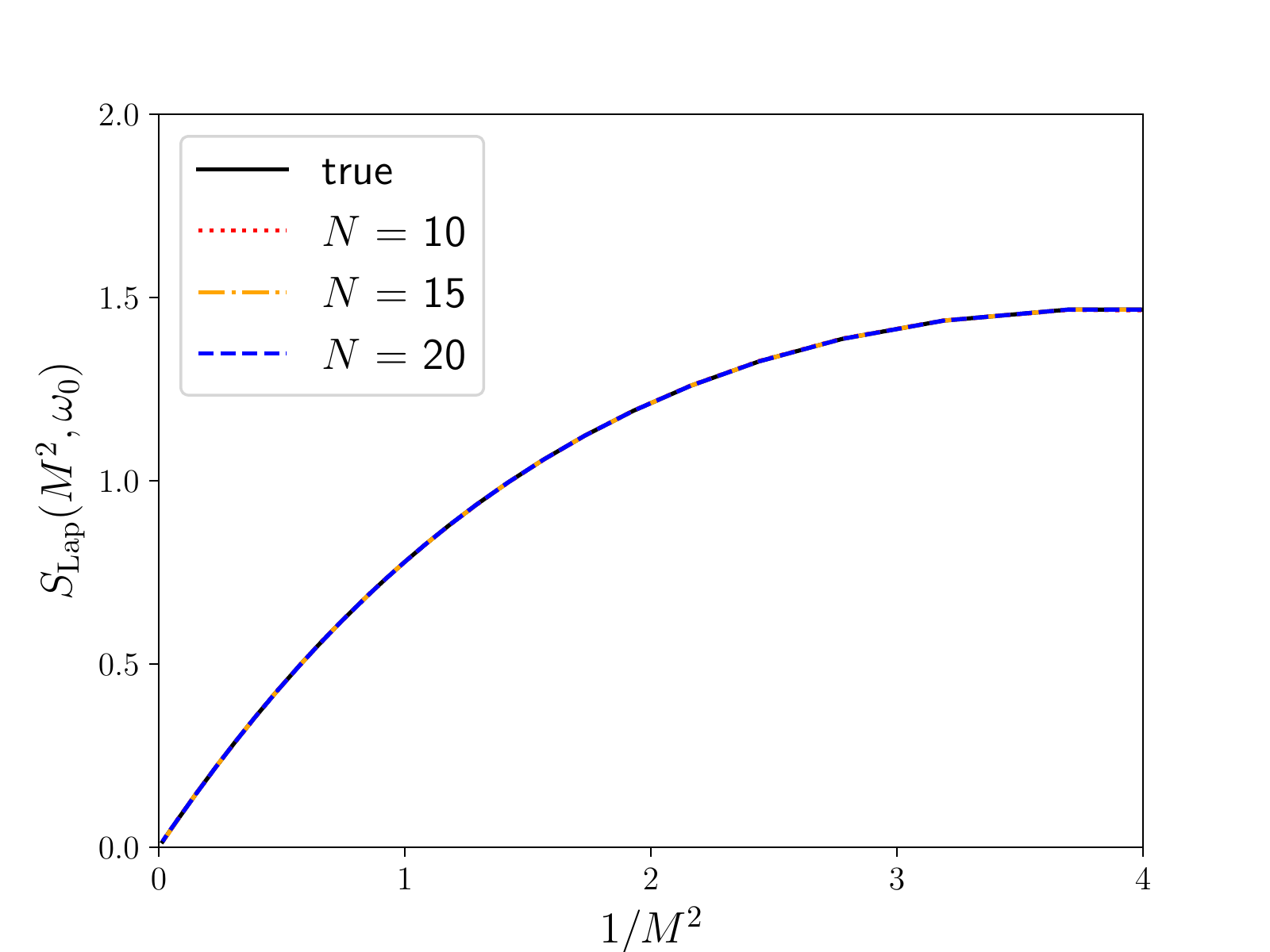}
  \caption{Chebyshev approximation of the kernel
    $S_{\rm Lap}(M^2,\omega_0)$ corresponding to the Laplace transform
    as a function of $1/M^2$.
    The true function (solid line) as well as its approximations
    (dotted, dot-dashed, dashed) are shown for $\omega_0$ = 1.0 (left)
    and 0.5 (right).
  }
  \label{fig:cheb_approx_exp}
\end{figure}

As another test, we consider the Laplace transform (\ref{eq:Laplace}),
which is achieved by a smearing kernel
\begin{equation}
  \label{eq:Laplace_smearing}
  S_{\rm Lap}(M^2,\omega') = \frac{2\omega'}{M^2} e^{-\omega'^2/M^2}.
\end{equation}
In Fig.~\ref{fig:cheb_approx_exp} we draw a curve of
$S_{\rm Lap}(M^2,\omega_0)$, which represents a
convolution with a trivial spectrum $\delta(\omega'-\omega_0)$,
together with its approximations (\ref{eq:rhoDelta_approx}).
The plots for $\omega_0$ = 1.0 (left) and 0.5 (right) demonstrate that
the Chebyshev polynomials provide a very precise approximation even
with $N$ = 10.
This is not unreasonable because the Laplace transform is a smooth
function over the entire range of energy.

\section{Smeared spectral function from lattice data:
  Charmonium correlators}
\label{sec:charmonium}
We test the method with LQCD data for Charmonium correlators.
Our data are obtained on the lattice with 2+1 flavors of M\"obius
domain-wall fermions.
They have been previously used for an extraction of the charm quark
mass from the Charmonium temporal moments \cite{Nakayama:2016atf}.
(The same lattice ensembles are also used for
calculations of the Dirac spectrum \cite{Cossu:2016eqs,Nakayama:2018ubk},
short-distance current correlators \cite{Tomii:2017cbt},
topological susceptibility \cite{Aoki:2017paw},
$\eta'$ meson mass \cite{Fukaya:2015ara}.)
Among 15 ensembles generated at various lattice spacings and lattice
sizes, we use the one at a lattice spacing $a$ = 0.080~fm, a lattice
size $32^3\times 64$, and bare
quark masses $am_{ud}$ = 0.007 and $am_s$ = 0.040.
The corresponding pion mass is 309(1)~MeV.
We take 100 gauge configurations and calculate the Charmonium
correlator with a tuned charm quark mass $am_c$ = 0.44037, and compute
charm quark propagators from $Z_2$ noises distributed over a time
slice to construct Charmonium correlators.
We then construct the Charmonium correlators, which correspond to
those of local currents in the pseudo-scalar (PP) and vector (VV)
channels.
This calculation has been repeated for 8 source time slices to
improve statistical signal, so that the total number of measurements
is 800.
Three spatial polarizations are averaged for the VV channel.

\begin{figure}[tbp]
  \centering
  \includegraphics[width=10cm]{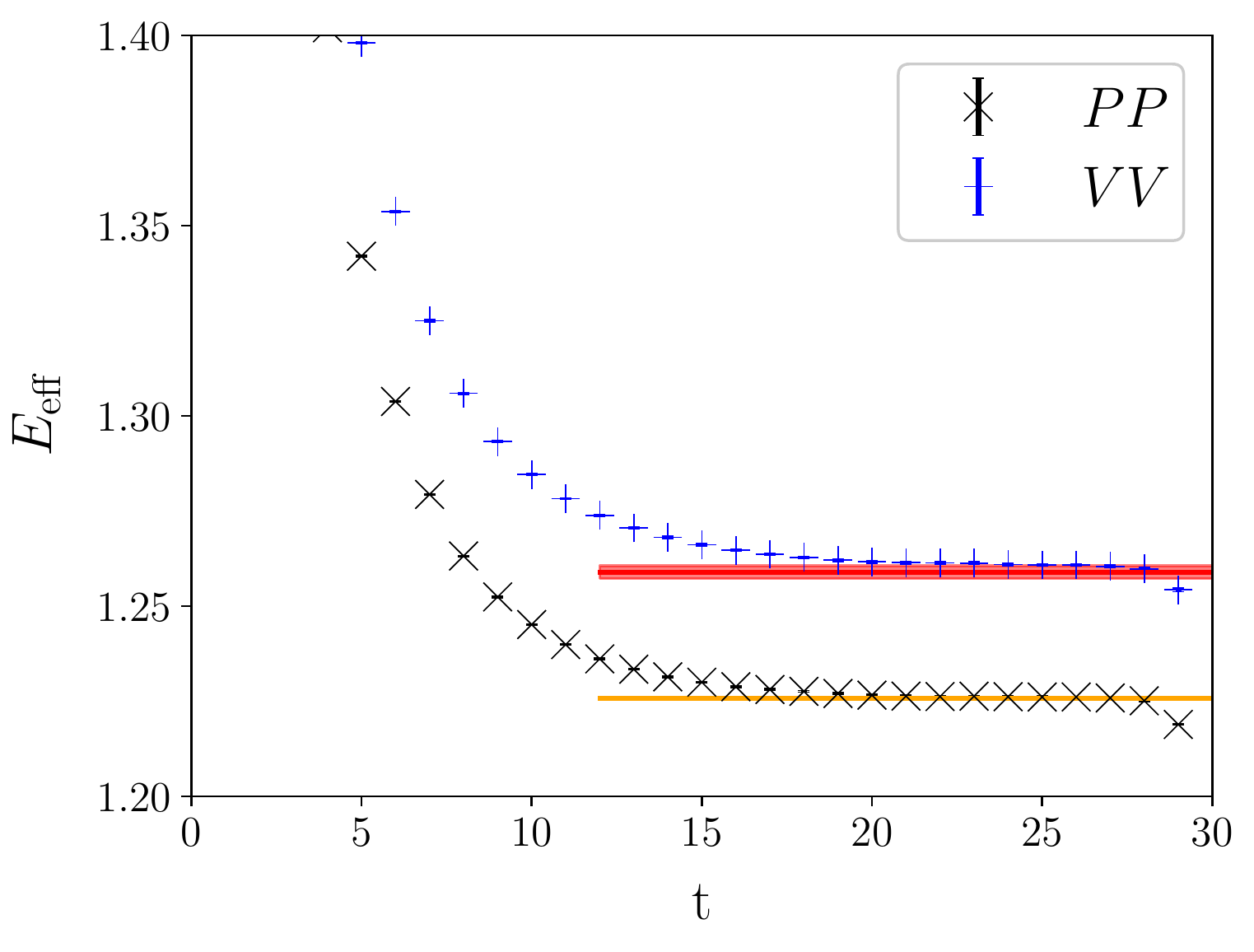}
  \caption{
    Effective mass of the pseudo-scalar (circles) and vector
    (squares) correlators calculated on the lattice with lattice
    cutoff $1/a$ = 2.453(4)~GeV.
    The charm quark mass is tuned such that the spin-averaged 1S mass
    matches the experimental data.
  }
  \label{fig:effmass}
\end{figure}

Fig.~\ref{fig:effmass} shows the effective mass
$E_{\rm eff}(t)=\ln[C(t)/C(t+1)]$.
We observe that the correlator is nearly saturated by the ground
state at around $t=18$, and the information of the excited states are
encoded in the region of smaller $t$'s.
Energy levels and amplitudes obtained by a multi-exponential fit of
the form $C(t)=\sum_iA_ie^{-E_it}$ are given in
Table~\ref{tab:mass-amp}.
We include four levels in the fit, yet show the results only up to
three levels since the error of the third amplitude is already large. 
Statistical correlation among different $t$'s is taken into account in
the fit.
The results reproduce the experimental data reasonably well.
For instance, the lowest-lying vector meson masses are 3.09 and
3.76(19)~GeV, which correspond to the experimentally observed $J/\psi$ and
$\psi(2S)$ states of masses 3.10 and 3.69~GeV, respectively.

\begin{table}[tbp]
  \centering
  \begin{tabular}{c|cc|cc}
    \hline
    & \multicolumn{2}{c|}{PP channel} & \multicolumn{2}{c}{VV channel}
    \\\hline
    $i$ & $E_i $ & $A_i$ & $E_i$ & $A_i$
    \\\hline
    0 & 1.22576(18) & 0.17638(31) & 1.2589(17) & 0.1308(35) \\
    1 & 1.521(19) & 0.205(22) & 1.534(79) & 0.184(62) \\
    2 & 1.831(15) & 0.25(17) & 1.834(15) & 0.04(88) \\
    \hline
  \end{tabular}
  \caption{Energy levels and amplitudes for the Charmonium correlators.}
  \label{tab:mass-amp}
\end{table}

\begin{figure}[tbp]
  \centering
  \includegraphics[width=10cm]{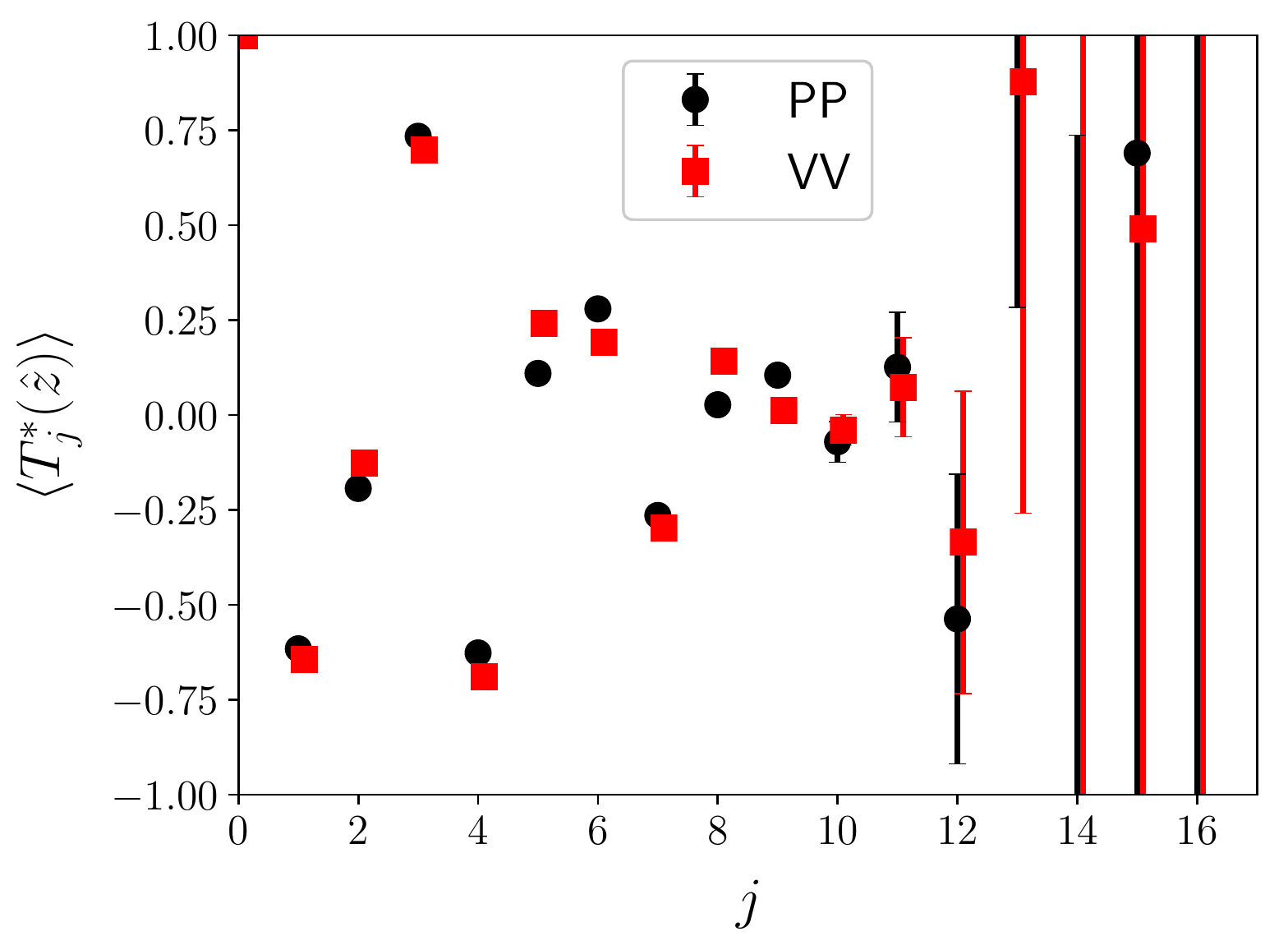}
  \caption{
    Chebyshev matrix elements $\langle T_j^*(\hat{z})\rangle$
    for the PP (circles) and VV (squares) Charmonium correlators.
    The VV points are slightly shifted horizontally for clarity.
    The statistical error is estimated using the jackknife method.
  }
  \label{fig:Chebz}
\end{figure}

We construct the Chebyshev matrix elements
$\langle T_j^*(\hat{z})\rangle$ defined in (\ref{eq:rhoDelta_approx})
from the current correlators.
The calculation is straightforward.
Namely, we replace the term of $\hat{z}^t$ in
the polynomials by $C(t+2t_0)/C(2t_0)$ and calculate the linear
combinations of them with the coefficients of the shifted Chebyshev
polynomials. 
We take $t_0=1$ in the lattice unit.
The results are shown in Fig.~\ref{fig:Chebz};
the error is calculated using the jackknife method.
It turned out that the Chebyshev matrix elements are precisely
determined up to $j=11$, which corresponds to $t=13$.
Beyond that point, the statistical error grows rapidly, and the
results eventually get out of the range of $\pm 1$, which must be
satisfied for the Chebyshev polynomials.

\begin{figure}[tbp]
  \centering
  \includegraphics[width=10cm]{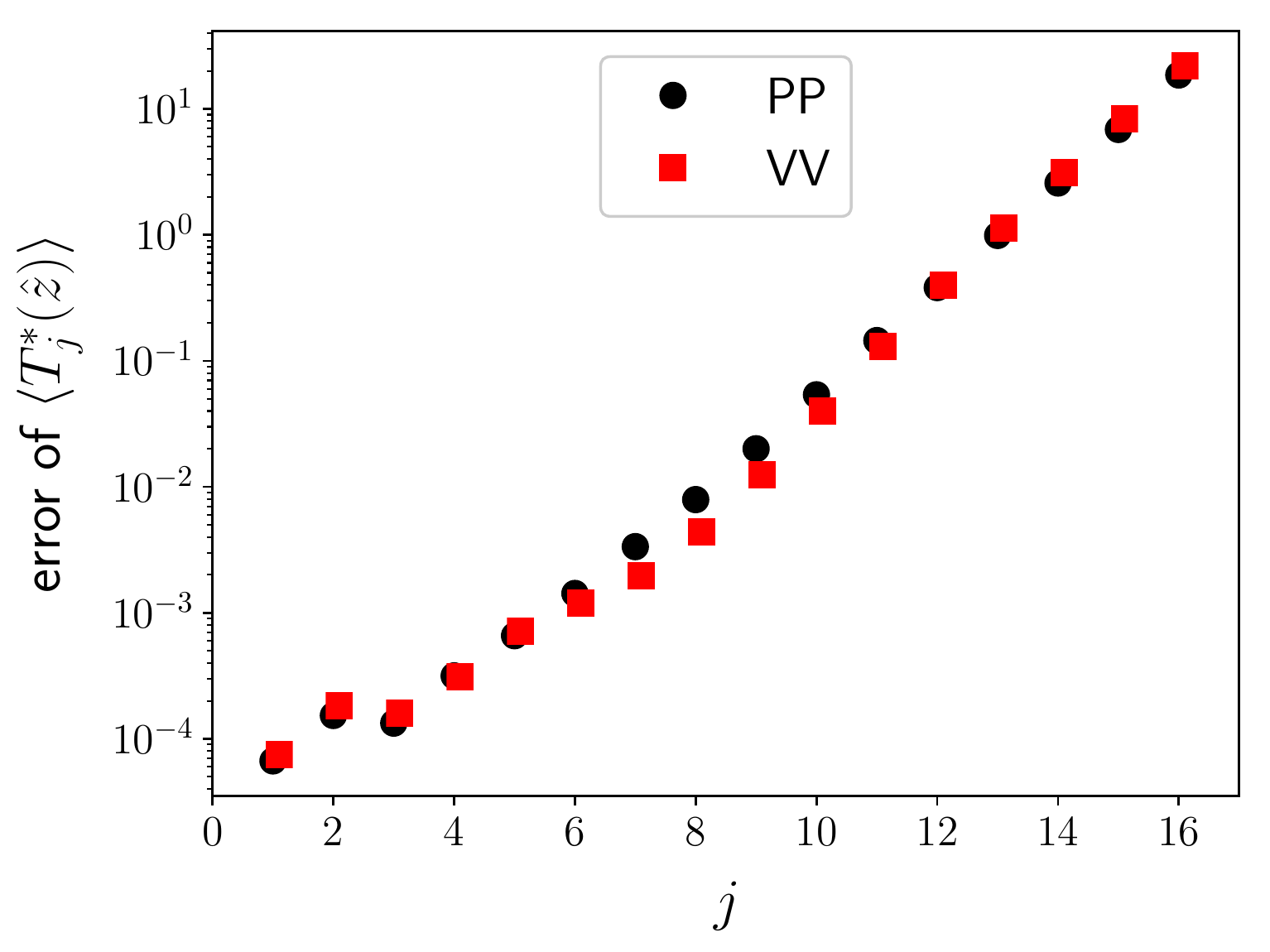}
  \caption{
    Statistical error of $\langle T_j^*(\hat{z})\rangle$ as a function
    of $j$.
    Circles and squares represent the data for PP and VV channels,
    respectively. 
  }
  \label{fig:error_Chebz}
\end{figure}

In fact, the statistical error grows exponentially for higher-order 
polynomials as shown in Fig.~\ref{fig:error_Chebz}.
Its growth rate is about a factor of three to proceed by another order
in $j$, which means that 10 times larger statistical samples would be
needed to include yet another order to improve the Chebyshev
approximation.
This is not unreasonable because we try to construct the quantities of
$O(1)$ as a linear combination of terms of exponentially different
orders. 
For the Charmonium correlator at the lattice spacing chosen in this
work, the terms of $\hat{z}^t$ are suppressed roughly by $e^{-1.2t}$,
which is however compensated by the Chebyshev coefficients growing
even faster.
In the end, a strong cancellation among different powers of $\hat{z}$
gives a number between $-1$ and $+1$, and the noise is relatively 
enhanced. 
For instance, at the order $j=12$, a cancellation of four orders of
magnitudes takes place and it becomes even harder for higher orders.

Since the Chebyshev approximation drastically fails outside
the domain $0\le x\le 1$, we are not able to use it beyond the order
where 
$|\langle T_j^*(\hat{z})\rangle|$
exceeds 1.
Instead, we introduce a fit to determine these matrix elements
$\bar{T}_j\equiv\langle T_j^*(\hat{z})\rangle$
in such a way that they are consistent with $\bar{C}(t)$ while
satisfying a constraint $|\bar{T}_j|\le 1$.
To do so, we can use the reverse formula of the shifted Chebyshev
polynomials \cite{Cheb_rev}:
\begin{equation}
  \label{eq:reverse_Cheb}
  x^n = 2^{1-2n}\sideset{}{'}\sum_{r=0}^n \left(
    \begin{array}[c]{c}
      2n\\ n-r
    \end{array}
  \right)
  T_r^*(x),
\end{equation}
where the prime on the sum indicates that the term of $r=0$ is to be
halved.
The relation to be satisfied is then
\begin{equation}
  \label{eq:to_fit}
  \bar{C}(t) = 2^{1-2t}
  \left[
    \frac{1}{2}\left(\begin{array}[c]{c} 2t\\t\end{array}\right)
    + \sum_{r=1}^t\left(
      \begin{array}[c]{c} 2t\\t-r\end{array}
    \right)\bar{T}_r
  \right].
\end{equation}
We take $T_r^*$'s as free parameters to be determined and fit the
lattice data $\bar{C}(t)$ with a constraint $|\bar{T}_j|\le 1$.
Statistical correlations of $C(t)$ among different $t$'s are taken
into account in the fit using the least-squares fit package
{\sf lsqfit} by Lepage \cite{lsqfit}.
The resulting values of $\bar{T}_j$ are listed in Table~\ref{tab:Tr}.
Since the numbers of inputs, $\bar{C}(t)$'s, and unknowns, $\bar{T}_j$'s,
are the same, the condition (\ref{eq:to_fit}) may be solved as a system
of linear equations unless the constraints are introduced.
In fact, the results are unchanged from the direct determination
through (\ref{eq:construct_Tj}) within the statistical error
for small $j$'s up to $j\simeq 10$.
Beyond that, they are affected by the constraints.
The error becomes large, of order 1, for large $j$'s, where they are
essentially undetermined by the fit but still kept within $\pm 1$.

\begin{table}[tbp]
  \centering
  \begin{tabular}[c]{cD{.}{.}{8}D{.}{.}{8}}
    \hline
    $j$ & \multicolumn{1}{c}{PP channel} & \multicolumn{1}{c}{VV channel}\\
    \hline
    1 & -0.6157(17) & -0.6447(17)\\
    2 & -0.1933(48) & -0.1257(49)\\
    3 &  0.7341(61) &  0.6980(63)\\
    4 & -0.6262(46) & -0.6892(49)\\
    5 &  0.1090(35) &  0.2410(43)\\
    6 &  0.2803(72) &  0.1914(78)\\
    7 & -0.267(17) & -0.297(17)\\
    8 &  0.036(38) &  0.144(37)\\
    9 &  0.073(81) & -0.001(78)\\
    10 &  0.03(16) &   0.01(15)\\
    11 & -0.14(26) & -0.08(25)\\
    12 &  0.08(36) &  0.07(35)\\
    13 &  0.06(43) &  0.03(42)\\
    14 & -0.07(51) & -0.05(51)\\
    15 & -0.05(60) & -0.02(59)\\
    16 & -0.01(77) &  0.00(76)\\
    \hline
  \end{tabular}
  \caption{Fit results for $\bar{T}_j$.}
  \label{tab:Tr}
\end{table}

\begin{figure}[tbp]
  \centering
  \includegraphics[width=8cm]{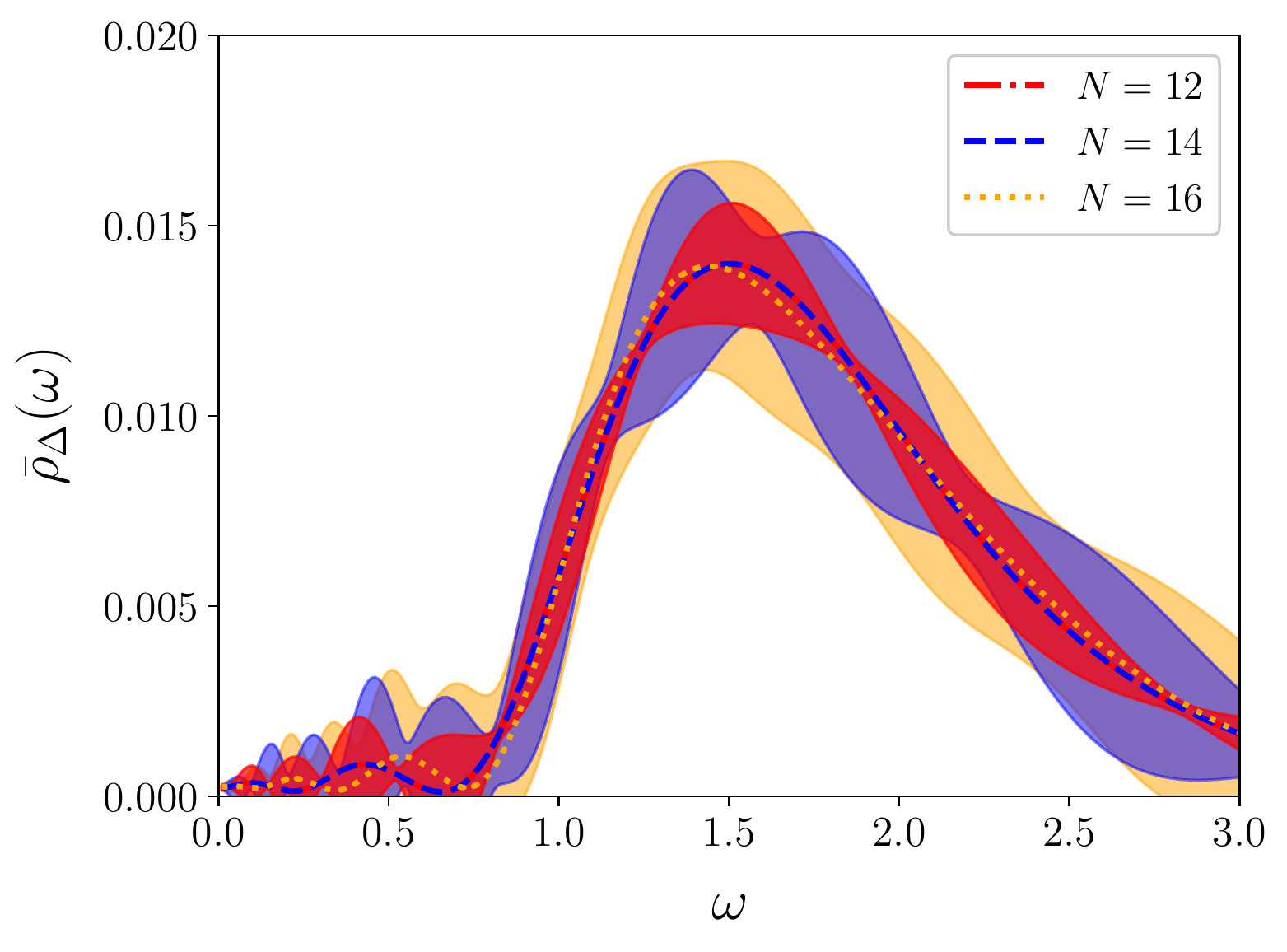}
  \includegraphics[width=8cm]{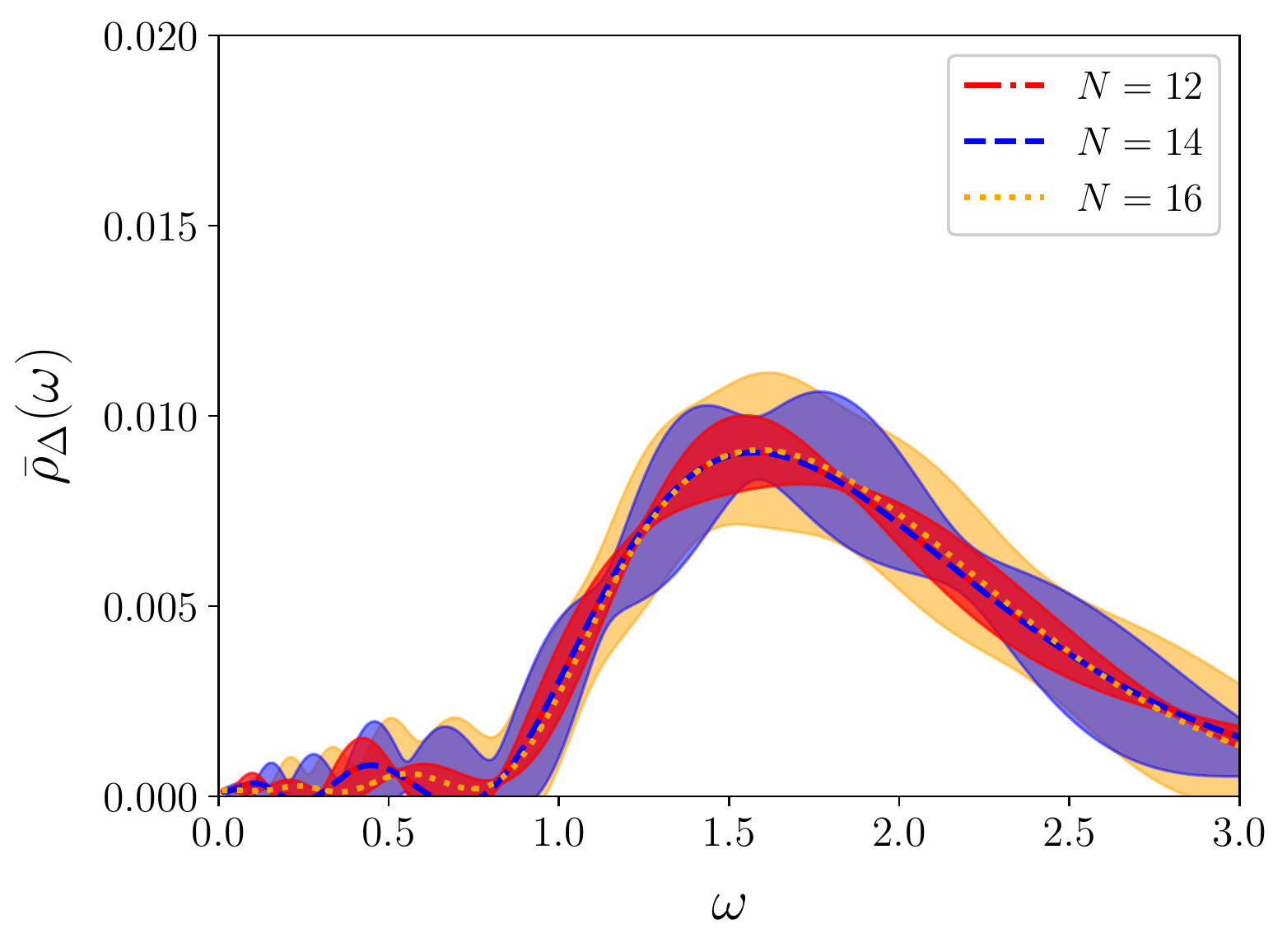}
  \includegraphics[width=8cm]{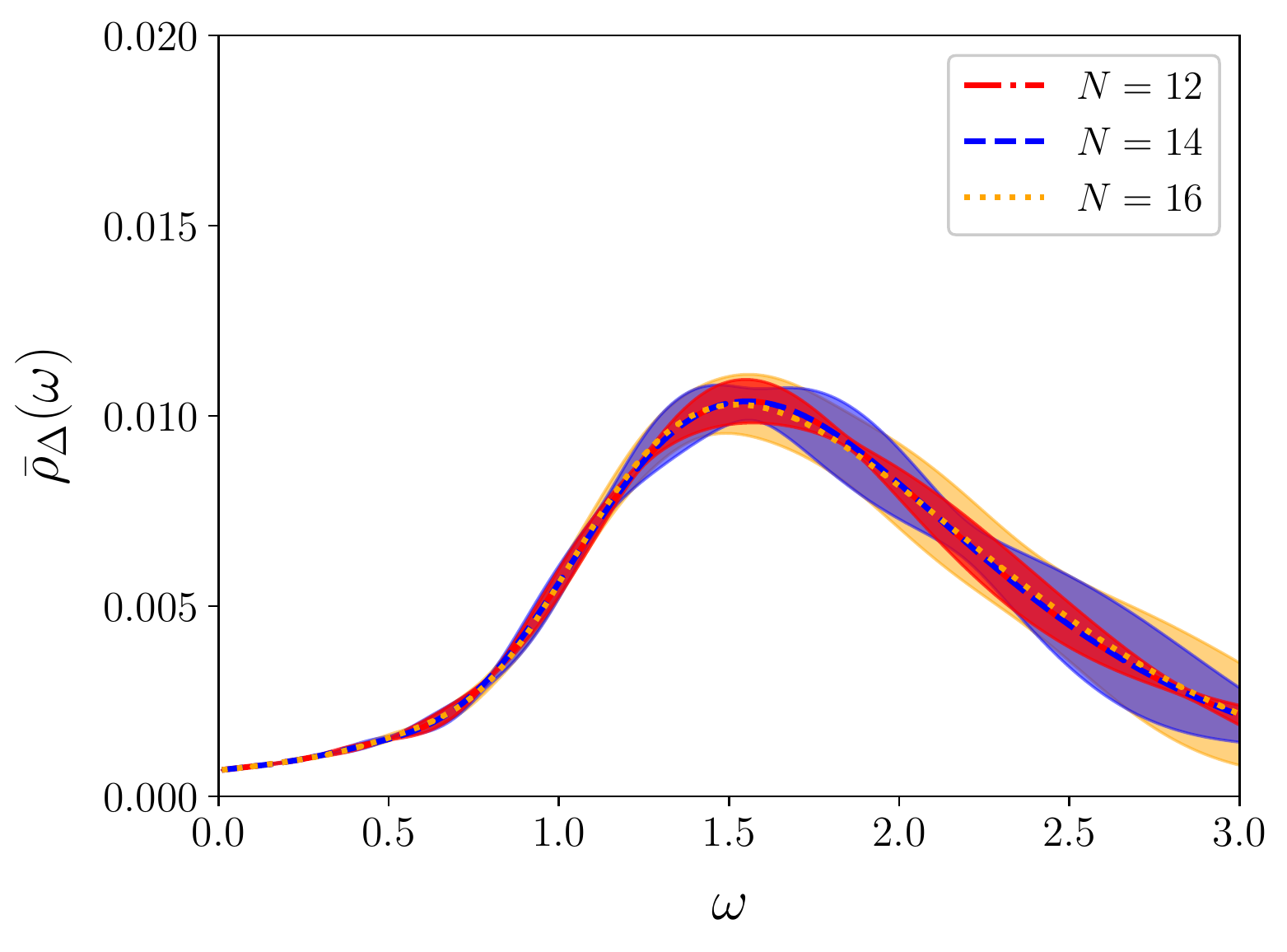}
  \includegraphics[width=8cm]{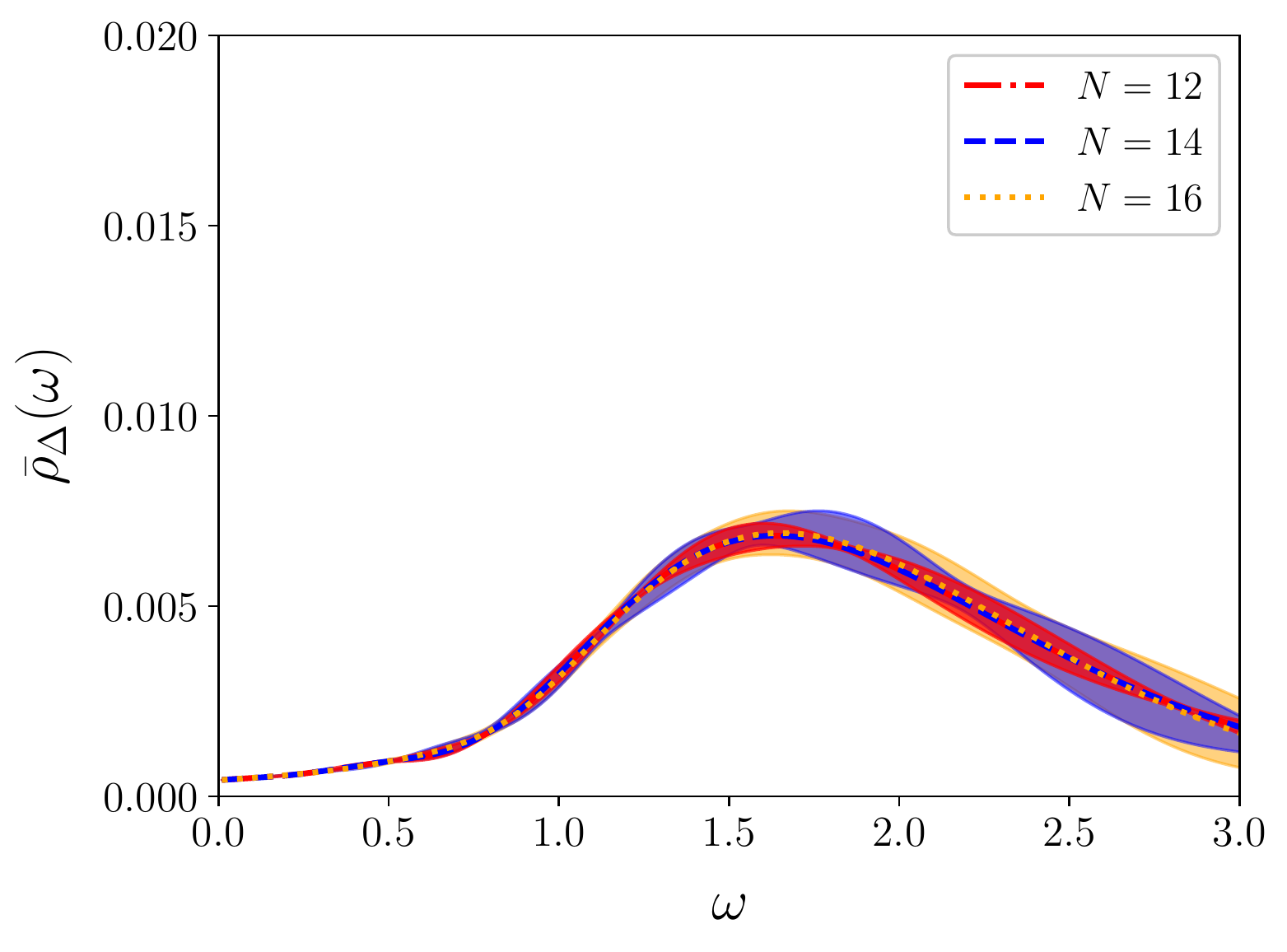}
  \caption{
    Smeared spectral function $\bar{\rho}_\Delta(\omega)$
    reconstructed using (\ref{eq:rhoDelta_approx}).
    The order of approximation is $N$ =
    12 (red), 14 (blue), 16 (orange).
    The results for $\Delta$ = 0.1 (top panels) and
    0.3 (bottom panels),
    for the PP channel (left panels) and VV channel (right panels),
    are shown. 
  }
  \label{fig:rho_variousN}
\end{figure}

Once a set of estimates for $\langle T_j^*(\hat{z})\rangle$,
{\it i.e.} $\bar{T}_j$,
is obtained, the remaining task is to use (\ref{eq:rhoDelta_approx})
to estimate $\bar{\rho}_\Delta(\omega)$.
The results are shown in Fig.~\ref{fig:rho_variousN}.
Three bands corresponding to the polynomial order $N$ = 12 (red), 14
(blue), 16 (orange) are overlaid.
We find that the overall shape is unchanged by adding more terms,
{\it i.e.} from 12 to 14 or to 16,
while the size and shape of the statistical error is affected.
When the polynomial order is lower, some wiggle structure is observed,
while the necks, the positions of small statistical error, are fatten
by adding more terms and eventually the error becomes nearly uniform
over $\omega$.
Beyond $N=16$, the results are essentially unchanged, since the higher 
order coefficients $c_j^*(\omega)$ are exponentially suppressed.

\begin{figure}[tbp]
  \centering
  \includegraphics[width=8cm]{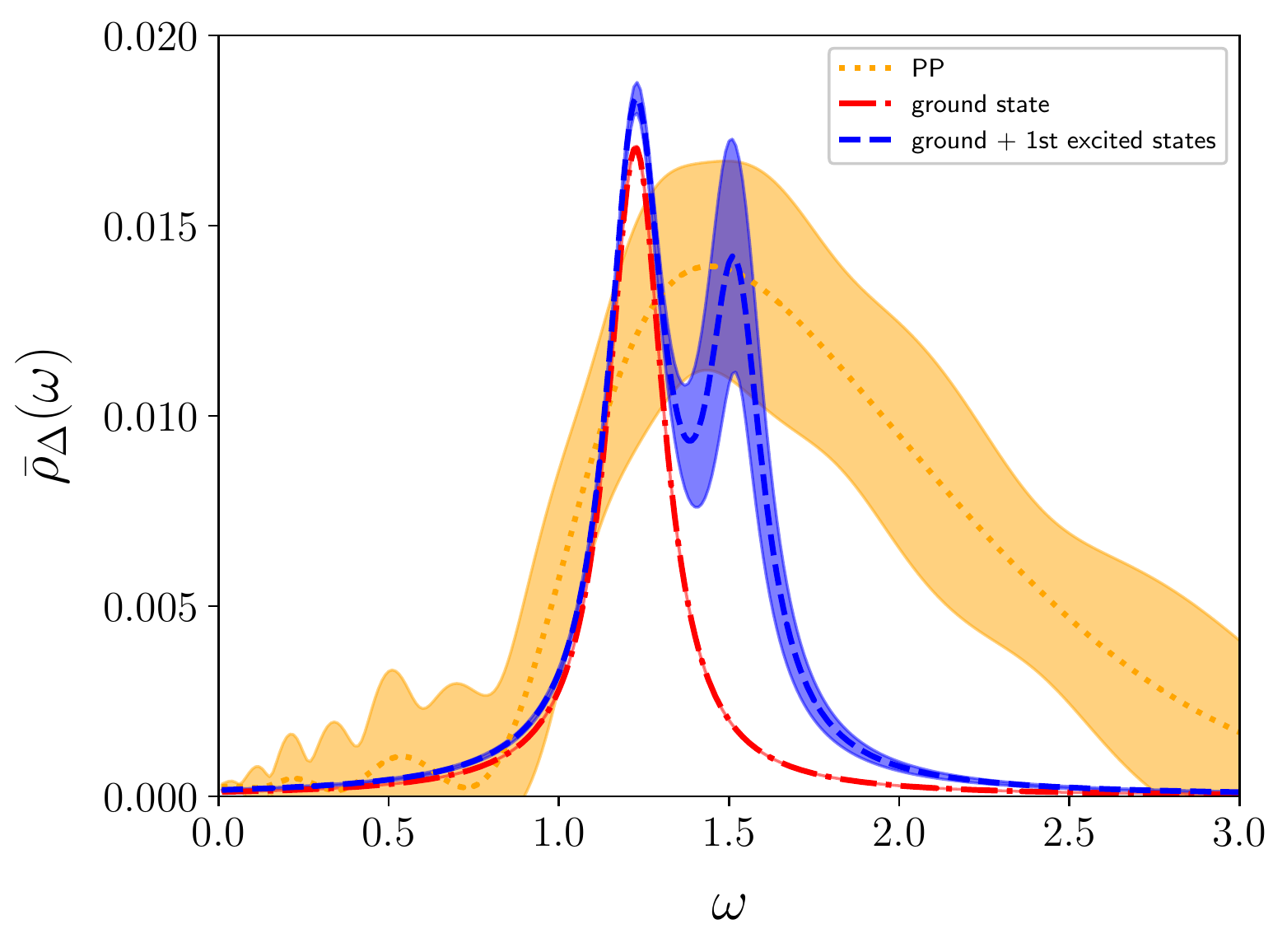}
  \includegraphics[width=8cm]{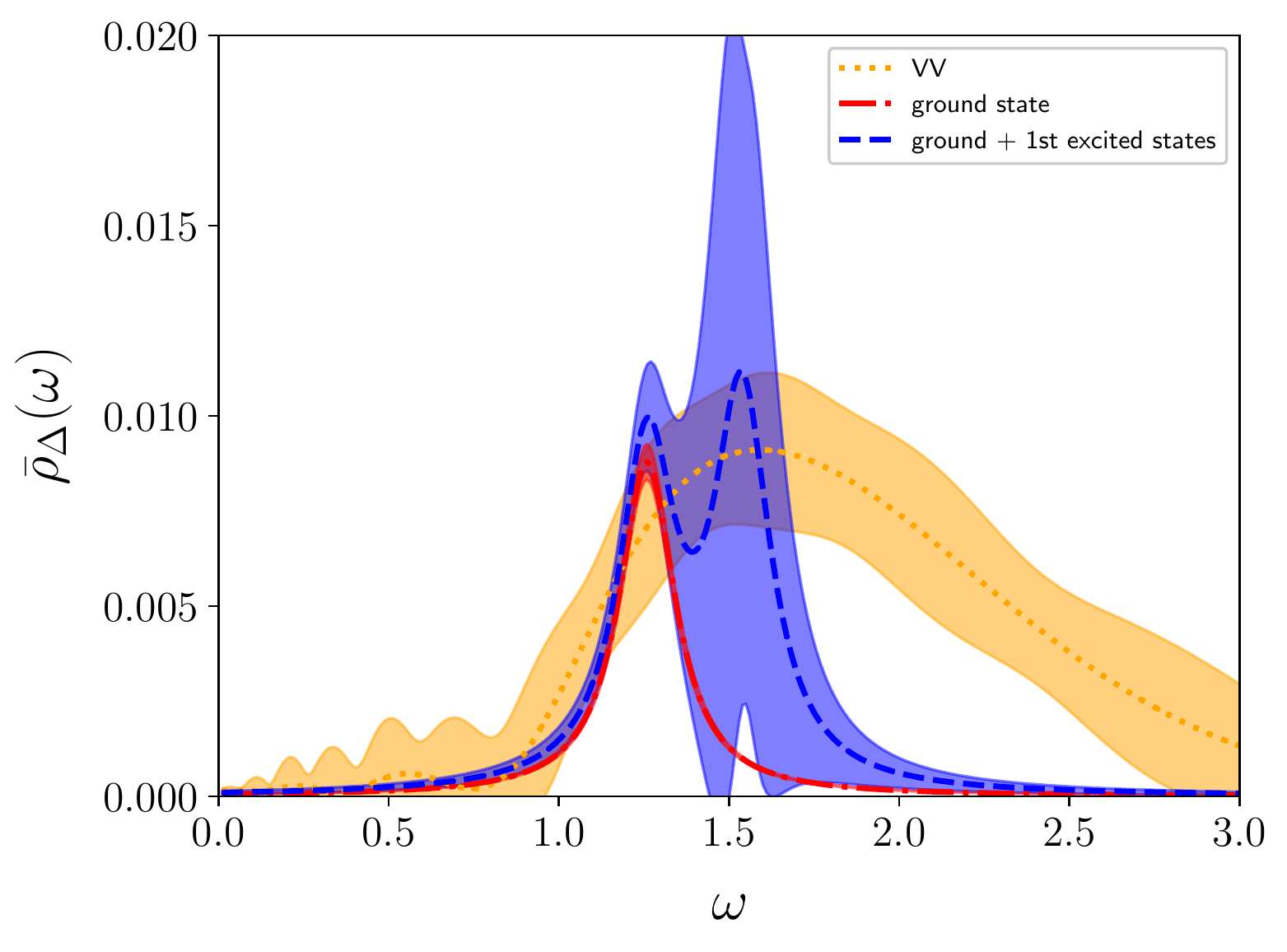}
  \includegraphics[width=8cm]{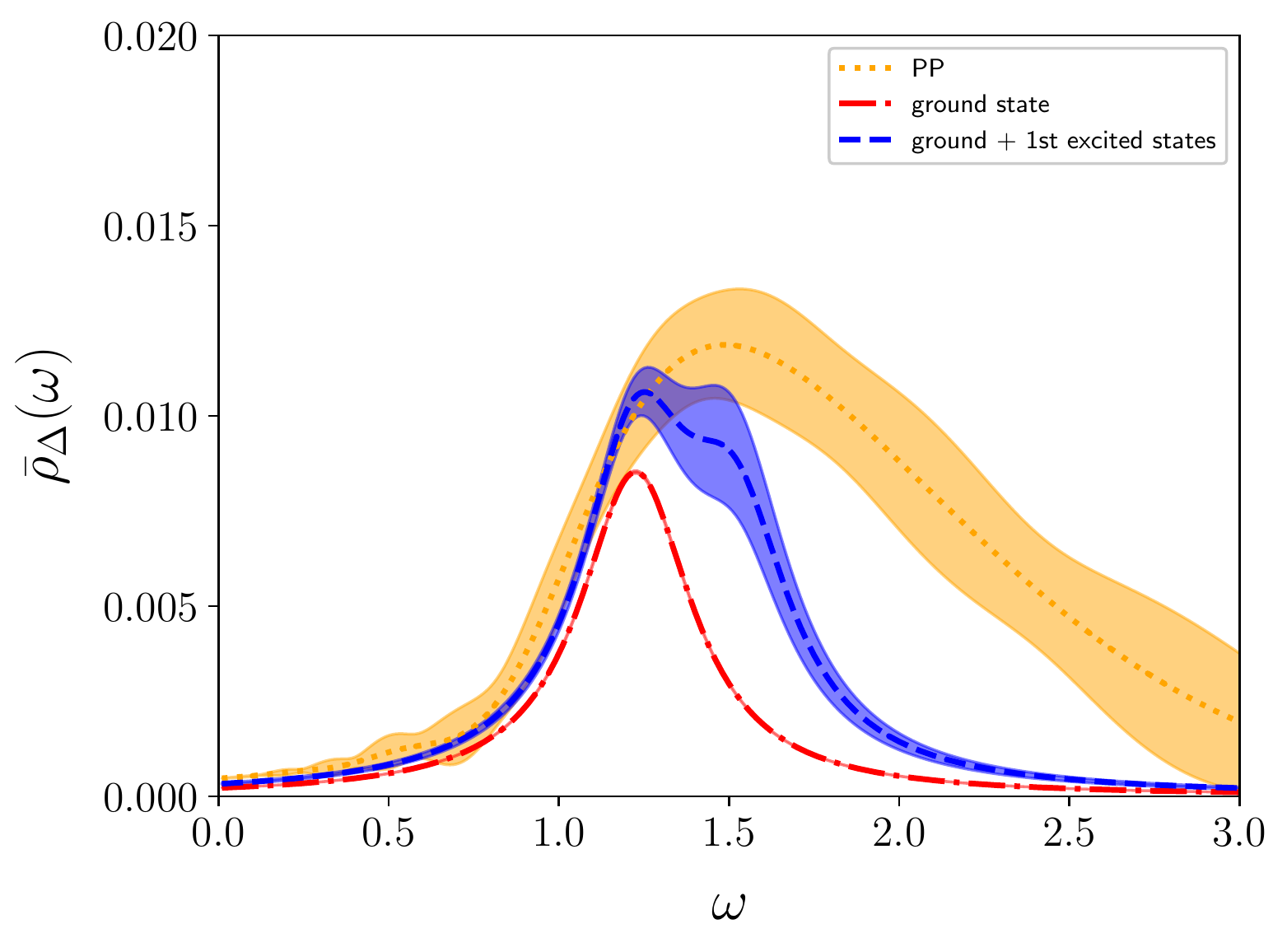}
  \includegraphics[width=8cm]{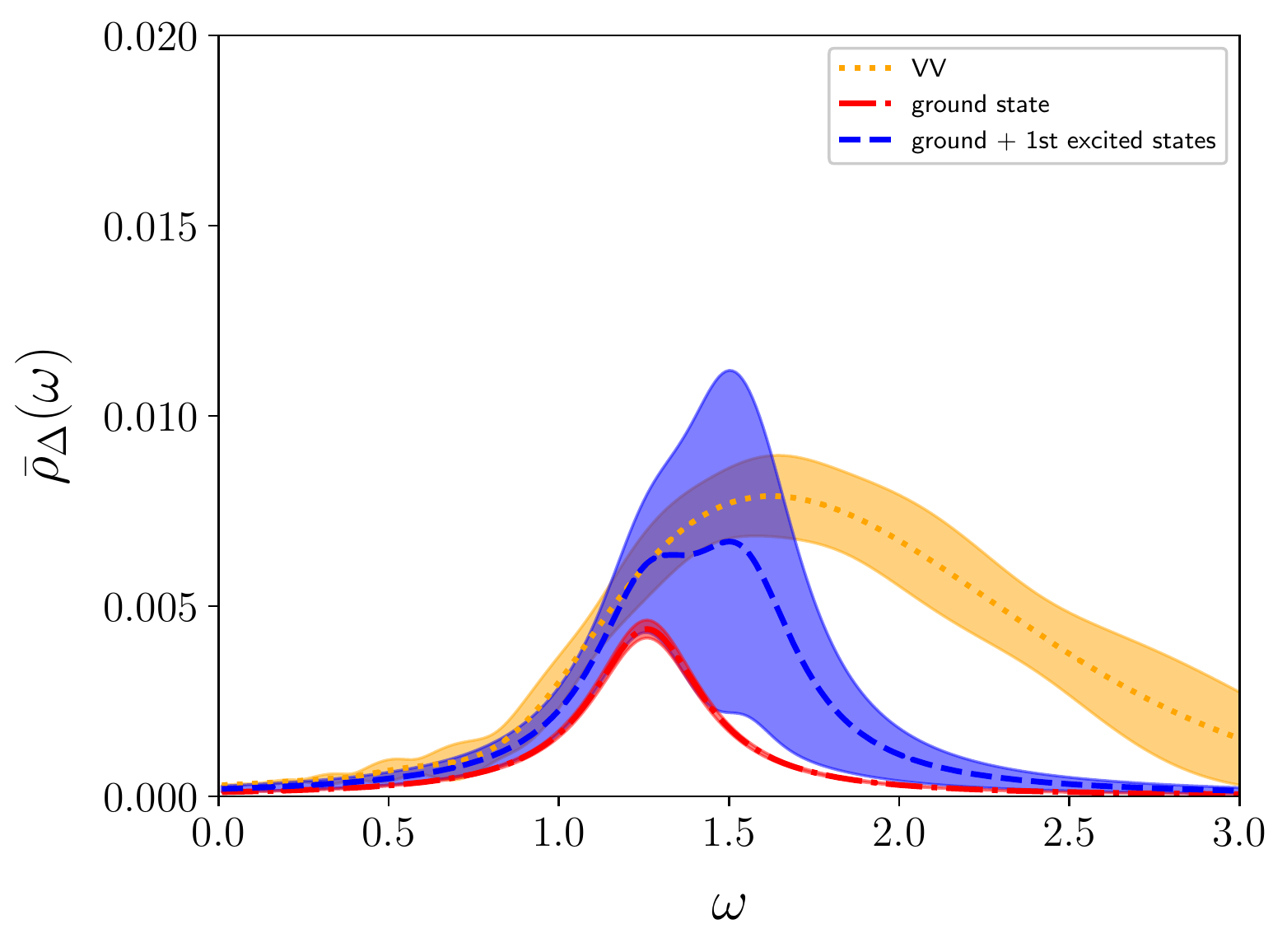}
  \includegraphics[width=8cm]{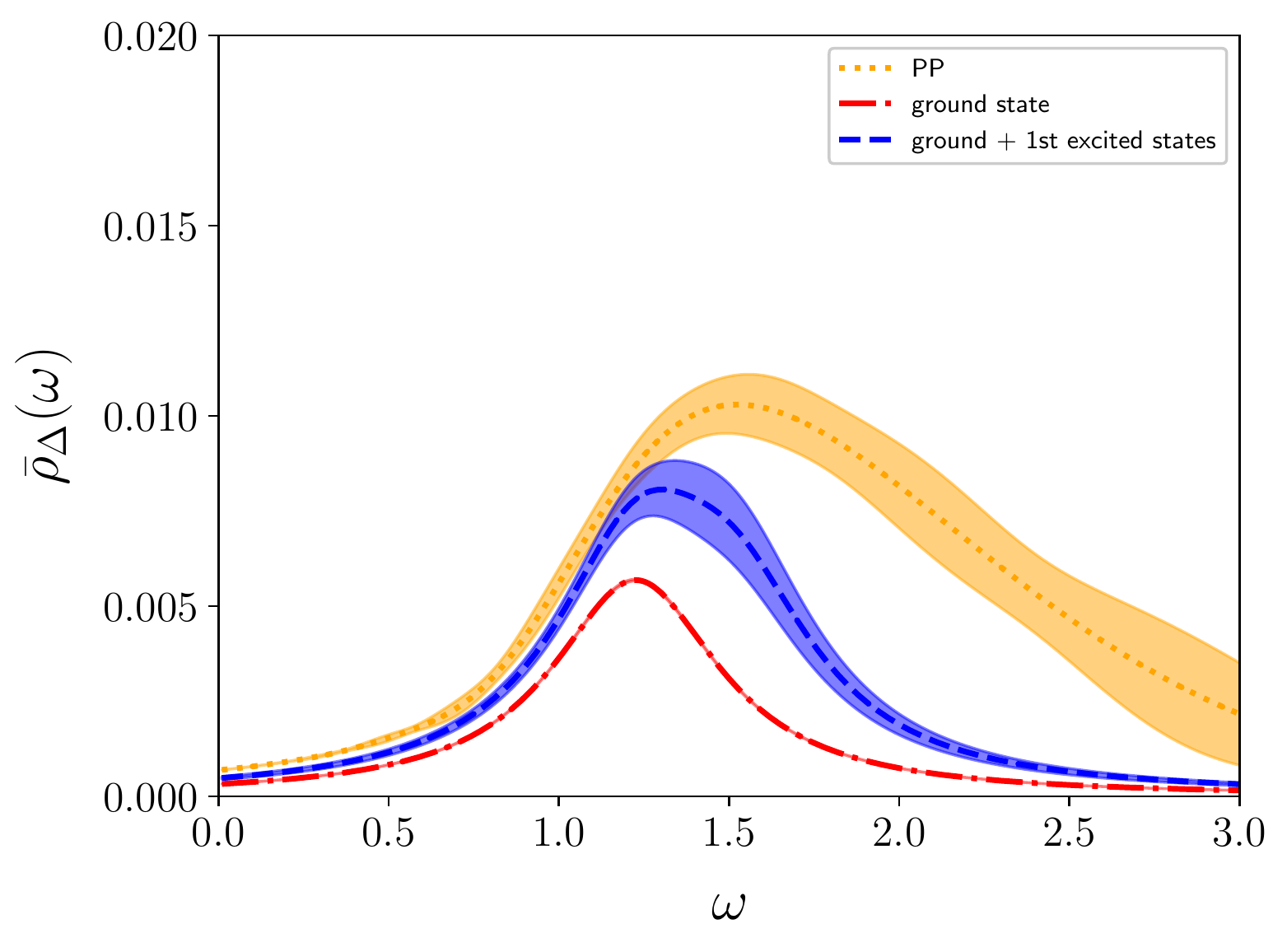}
  \includegraphics[width=8cm]{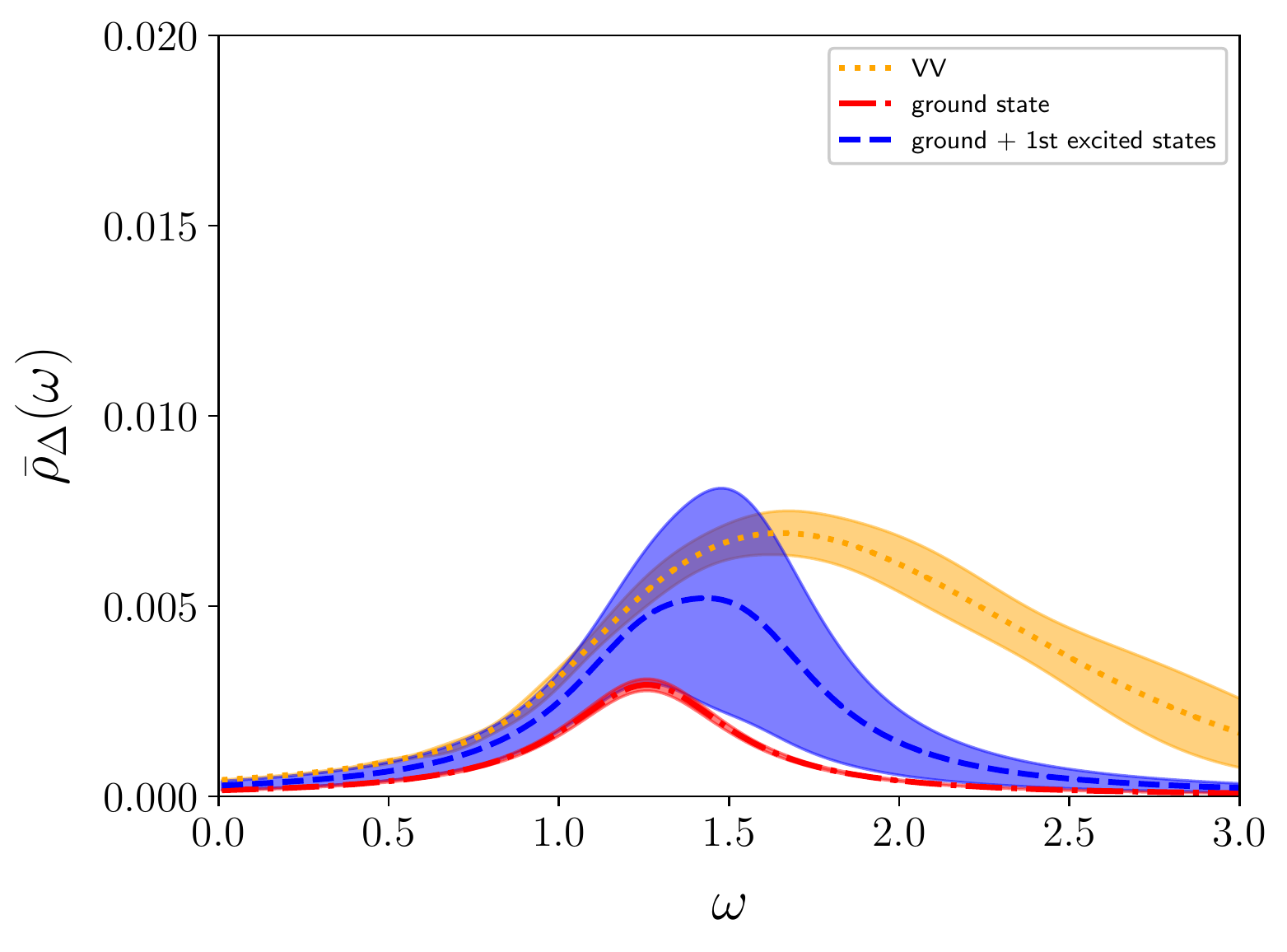}
  \caption{
    Reconstructed smeared spectral function $\rho_\Delta(\omega)$ for
    $\Delta$ = 0.1 (top), 0.2 (middle) and 0.3 (bottom).
    Left and right columns are those of the pseudo-scalar and vector
    channels. 
    Results are shown by orange band, while the 
    ground-state contribution (dot-dashed) and
    the ground-state and plus excited-state contribution (dashed) are
    plotted assuming they have $\delta$-function structures.
  }
  \label{fig:rhoDelta}
\end{figure}

In Fig.~\ref{fig:rhoDelta} we show the results for the smeared
spectral function $\bar{\rho}_\Delta(\omega)$ obtained with $N=16$.
The smearing kernel is $S_\Delta(\omega,\omega')$ with $\Delta$ = 0.1
(top), 0.2 (middle) and 0.3 (bottom).
The results are compared with the expected contributions from the
ground state and first excited state (red and blue curves).
They are drawn assuming $\delta$-function distributions,
$\bar{\rho}(\omega')=\sum_i A_ie^{-\omega't_0}\delta(\omega'-E_i)$,
with the fitted values of energy levels $E_i$ and their amplitudes
$A_i$ given in Table~\ref{tab:mass-amp}.
The factor $e^{-\omega't_0}$ is introduced to take account of the time
evolution from 0 to $t_0$, which is included in the definition of
the state $|\psi\rangle=e^{-Ht_0}J_\mu|0\rangle$.

When the smearing width is large, $\Delta=0.3$ (bottom panels of Fig.~\ref{fig:rhoDelta}), we observe that the
reconstructed smeared spectral function follows the expected form from 
the low-lying states in the lower region of $\omega$.
As $\omega$ increases, the spectral function indicates more
contributions from higher excited states.
This is exactly what we expected.
The lattice data in the short time separations contain the
information of such states, which is properly extracted with our
method. 
From perturbation theory, one expects a constant proportional to the
number of color degrees of freedom $N_c=3$ for the (unsmeared)
spectral function $\rho(\omega)$.
This constant is slightly distorted because of the difference between
$\rho(\omega)$ and $\bar{\rho}(\omega)$.
For $t_0=1$ (in the lattice unit), this should give an exponentially
decreasing spectral function $\bar{\rho}_\Delta(\omega)$ as
$\sim \omega^2e^{-2\omega}$ at large $\omega$, which is indeed observed
in the results. 
On the other hand, the resonance structure is smeared out and
invisible with $\Delta=0.3$ as one can see from the contributions of
the ground and first-excited states.

For smaller smearing widths, $\Delta$ = 0.2 and 0.1, a larger
systematic error is expected due to the truncation of the
Chebyshev approximation.
For $\Delta=0.2$, a typical size of the error is about 10--20\% as
one can see from Figures~\ref{fig:cheb_approx} and
\ref{fig:cheb_approx_w0=0.5}.
(Dot-dashed lines correspond to $N=15$.)
This error due to the truncation is not included in the band shown in
Fig.~\ref{fig:rhoDelta}, but taking account of this marginal size of
error the reconstructed $\rho_\Delta(\omega)$ looks reasonable
for $\Delta=0.2$ (middle panels).
Namely, it follows the expected curve of the ground and the first
excited states up to around the peak of the latter and then drops
slowly due to higher excited state contributions.

The truncation error increases to 50--100\% for $\Delta=0.1$
(upper panels of Figures~\ref{fig:cheb_approx} and
\ref{fig:cheb_approx_w0=0.5}), and we should not take the results
(top panels of Fig.~\ref{fig:rhoDelta}) too seriously.
If it were precisely calculated, we would be able to resolve the
resonance structures as the curve of the low-lying state 
contributions suggests.
To do so, we need to include higher order terms of the Chebyshev
approximation, which requires much better precision of the simulation
data.
This reflects the fact that the reconstruction of the full spectral
function from Euclidean lattice data is an ill-posed problem.
One needs ridiculously high precision in order to achieve a full
reconstruction as emphasized in \cite{Hansen:2019idp}.




\section{Discussions}
\label{sec:discussions}
As already mentioned earlier, our proposal to calculate the smeared
spectral function is not limited to the case just discussed.
Any sort of weighted integral of the spectral function can be
considered. 
A well-known example is the contribution of quark vacuum polarization
to the muon anomalous magnetic moment $g-2$.
Phenomenologically, one employs the optical theorem and dispersion
relation to relate the vacuum polarization function in the Euclidean
domain $\Pi(Q^2)$ to a weighted integral of the experimentally
observed $R$-ratio, or the spectral function.
Then, an integral of $\Pi(Q^2)$ with an appropriate weight gives
the contribution to $g-2$.
In this case, however, a direct expression in terms of the time
correlator $C(t)$ is known \cite{Bernecker:2011gh}, and we do not
really need the approximation method developed in this work.

Another phenomenologically interesting example is the hadronic $\tau$
decay.
Using the finite-energy sum rule \cite{Braaten:1991qm}
the hadronic width can be written as
\begin{equation}
  \label{eq:hadronic_tau}
  R_\tau^{ud}=12\pi^2 S_{\rm EW}|V_{ud}|^2
  \int_0^{m_\tau^2}\frac{ds}{m_\tau^2}
  \left(1-\frac{s}{m_\tau^2}\right)^2
  \left(1+\frac{2s}{m_\tau^2}\right)
  \rho_{V+A}(s),
\end{equation}
where $S_{\rm EW}$ is a short-distance electroweak correction and
$|V_{ud}|$ is a CKM matrix element.
(Here, only the $ud$ contribution is considered. An extension to the
$us$ contribution is straightforward.)
The spectral function $\rho_{V+A}(s)$ denotes a sum of $VV$ and $AA$
channels. 
The integral (\ref{eq:hadronic_tau}) reflects a particular kinematics
of $\tau$ decay and has a complicated form, but our method can be
applied for such a case in principle.
A practical question would be, however, whether a good enough approximation
can be achieved with a limited number of terms.

The Laplace transform (\ref{eq:Laplace_smearing}) is often
considered in QCD sum rule analyses \cite{Shifman:1978bx}
because the corresponding Borel transform of perturbative series makes
it more convergent.
Our method allows to calculate two-point function after the Borel
transform directly using lattice QCD.
It may be useful to test the perturbative expansion and the
operator product expansion involved in the QCD sum rule calculations. 
Conversely, it can also be used to validate lattice QCD calculations
especially in the short-distance region.
Such a test has been performed using short-distance correlators in the
coordinate space \cite{Tomii:2017cbt}, and it is interesting to do the
test for the Borel transformed quantities.

The dispersion integral of the form (\ref{eq:dispersion}) is of course
another type of applications of our method.
Although the vacuum polarization function $\Pi(Q^2)$ in the space-like
momenta $Q^2$ can be directly obtained by a Fourier transform of the
lattice correlators, the Chebyshev approximation offers a method to
extract it at arbitrary values of $Q^2$ corresponding to the 
momenta of non-integer multiples of $2\pi/L$.
The Laplace transform as introduced in \cite{Feng:2013xsa} can do this 
too, but requires the information of large time separations in the
integral of the form $\int_0^\infty dt\,e^{\omega t}C(t)$.
The method developed in this work accesses only to relatively short
time separations, but we need to introduce an approximation.
Systematic error in each case has to be carefully examined.

Extension to the cases of more complicated quantities, such as the
nucleon structure function as measured in deep inelastic scattering or
the inclusive hadron decays, can also be considered.
One of the authors has proposed an analysis to use the dispersion
integral to relate the inclusive decay rate to an amplitude in
space-like momenta \cite{Hashimoto:2017wqo}.
This method contains a difficulty of requiring the information at
unphysical momentum regions, which may be avoided by more flexible
integral transformation proposed in this work.

For this class of applications, there are two independent kinematical
variables, $q^2$ and $p\cdot q$, with $p$ the momentum of a decaying
particle (or the initial nucleon) and $q$ the momentum transfer.
In order to apply the method outlined in this work, we need to fix one of
these kinematical variables and introduce a smearing on the other
variable.
More complicated integral might be useful for $b\to u\ell\nu$ decay
analysis for which one introduces elaborate kinematical cuts in order
to avoid backgrounds from $b\to c\ell\nu$.
The flexibility of our method would allow such analyses.

Going to high tempertature QCD, our method would not work as it is,
because the correlator can not be simply written as
$\langle\psi|\hat{z}^t|\psi\rangle$
due to the contribution from the opposite time direction.
The operator is then no longer a power of the transfer matrix, direct
estimate of the Chebyshev matrix elements is not available.

\section{Conclusions}
\label{sec:conclusions}
Precise reconstruction of the spectral function from lattice data
remains a difficult problem.
Instead, we calculate a smeared counterpart, which contains some
information of the spectral function after smearing out its detailed
structures.
For the Charmonium spectral function we obtain a reasonably precise
result for a smearing width $\Delta$ = 0.3, which is about 700~MeV in
the physical unit.
Since the mass splittings experimentally observed are narrower than
this, we are not able to resolve the details of the spectrum.
Taking the limit of small smearing width requires exponentially better
statistical precision, and it comes back to the original problem.
Still, our proposal has advantages compared to previously available
methods. 

In contrast to the Bayesian approach \cite{Burnier:2013nla} or the
maximum entropy method \cite{Nakahara:1999vy,Asakawa:2000tr},
our method allows a reliable estimate of the systematic errors, since
it does not assume any statistical distribution of unknown function.
In principle, the method is deterministic once the input lattice data
are given.
The least-squared fit involved in order to enforce the constraint
that eigenvalues of Hamiltonian is positive plays only a minor role
that becomes irrelevant when the lattice data are made precise.

Compared to the Backus-Gilbert method \cite{Hansen:2017mnd}, the
method proposed in this paper is more flexible as it allows any
predefined smearing function, while it is automatically determined in
the Backus-Gilbert method and therefore is uncontrollable.
The variant of the Backus-Gilbert method \cite{Hansen:2019idp} also
has this flexibility.
Our method also allows systematic improvements since the
approximation is achieved by a series of exponentially decreasing
coefficients. 
As the statistical precision of the input correlator is improved, one
can include higher order terms and thus improve the approximation.

The method can be used in the analysis of inclusive processes to
define intermediate quantities for which fully non-perturbative
lattice calculation is possible.
Its potential application is not limited to the spectral function for
two-point correlators, but other processes such as deep inelastic
scattering and inclusive $B$ meson decays can be considered.
Since the method does not rely on perturbation theory, the processes
with small momentum transfer can be calculated on a solid theoretical
ground, which was not available so far.
More importantly, we do not have to rely on the assumption of
quark-hadron duality.

\section*{Acknowledgments}
We thank the members of the JLQCD collaboration for discussions and
for providing the computational framework and lattice data.
Numerical calculations are performed 
on Oakforest-PACS supercomputer operated by Joint Center for Advanced
High Performance Computing (JCAHPC), as well as
on SX-Aurora TSUBASA at KEK used under its ``Particle, Nuclear, and Astro Physics Simulation Program.''
This work is supported in part by JSPS KAKENHI Grant Number 18H03710
and by the Post-K supercomputer project through the Joint Institute
for Computational Fundamental Science (JICFuS).


\end{document}